\documentclass[useAMS,usenatbib]{mn2e}

\usepackage{graphicx}
\usepackage{rotating}
\usepackage{amsmath}
\usepackage{txfonts}
\usepackage{color}
\usepackage{ulem}
\usepackage[hyphens]{url}
\usepackage{hyperref}

\title[A PSF-based approach to \textit{Kepler/K2} data. I.]{ A
  PSF-based approach to \textit{Kepler/K2} data. I. Variability within
  the \textit{K2} Campaign 0 star clusters M\,35 and
  NGC~2158.\thanks{Based on observation with the \textit{Kepler}
    telescope and with the Schmidt 67/92 cm Telescope at the
    Osservatorio Astronomico di Asiago, which is part of the
    Osservatorio Astronomico di Padova, Istituto Nazionale di
    AstroFisica.}
} 

\author[Libralato et al.]{ 
  M.\ Libralato\thanks{E-mail: \href{mailto:mattia.libralato@studenti.unipd.it}{mattia.libralato@studenti.unipd.it}}$^{1,2}$,
  L.\ R.\ Bedin$^{2}$, D.\ Nardiello$^{1,2}$, G.\ Piotto$^{1,2}$ \\
$^{1}$ Dipartimento\ di Fisica e Astronomia, Universit\`a di Padova, Vicolo dell'Osservatorio 3, Padova, I-35122, Italy \\
$^{2}$ INAF-Osservatorio Astronomico di Padova, Vicolo dell'Osservatorio 5, Padova, I-35122, Italy \\
}

\begin{document}

\date{Received 03 September 2015 / Accepted 05 November 2015}

\pagerange{\pageref{firstpage}--\pageref{lastpage}} \pubyear{2015}

\maketitle

\label{firstpage}

\begin{abstract}
\textit{Kepler} and \textit{K2} data analysis reported in the
literature is mostly based on aperture photometry. Because of
\textit{Kepler}'s large, undersampled pixels and the presence of
nearby sources, aperture photometry is not always the ideal way to
obtain high-precision photometry and, because of this, the data set
has not been fully exploited so far. We present a new method that
builds on our experience with undersampled \textit{HST} images. The
method involves a point-spread function (PSF) neighbour-subtraction
and was specifically developed to exploit the huge potential offered
by the \textit{K2} ``super-stamps'' covering the core of dense star
clusters. Our test-bed targets were the NGC~2158 and M\,35 regions
observed during the \textit{K2} Campaign 0. We present our PSF
modeling and demonstrate that, by using a high-angular-resolution
input star list from the Asiago Schmidt telescope as the basis for PSF
neighbour subtraction, we are able to reach magnitudes as faint as
$K_{\rm P}$$\simeq$24 with a photometric precision of 10\% over 6.5
hours, even in the densest regions. At the bright end, our photometric
precision reaches $\sim$30 parts-per-million. Our method leads to a
considerable level of improvement at the faint magnitudes ($K_{\rm
  P}\gtrsim$15.5) with respect to the classical aperture
photometry. This improvement is more significant in crowded
regions. We also extracted raw light curves of $\sim$60\,000 stars and
detrended them for systematic effects induced by spacecraft motion and
other artifacts that harms \textit{K2} photometric precision. We
present a list of 2133 variables.
\end{abstract}


\begin{figure*}
  \centering
  \includegraphics[width=\textwidth]{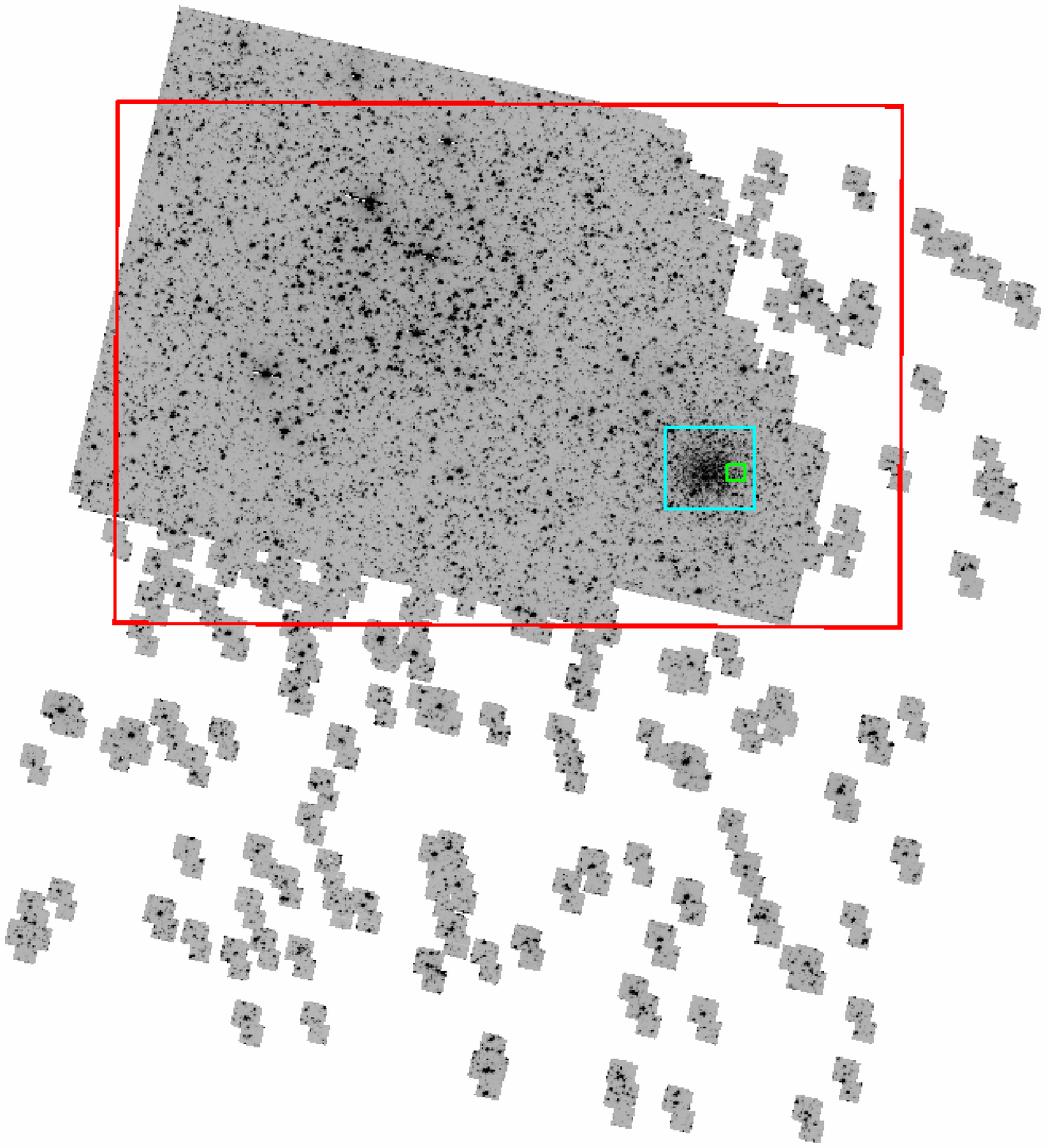}
  \vskip -35 pt
  \includegraphics[width=\textwidth]{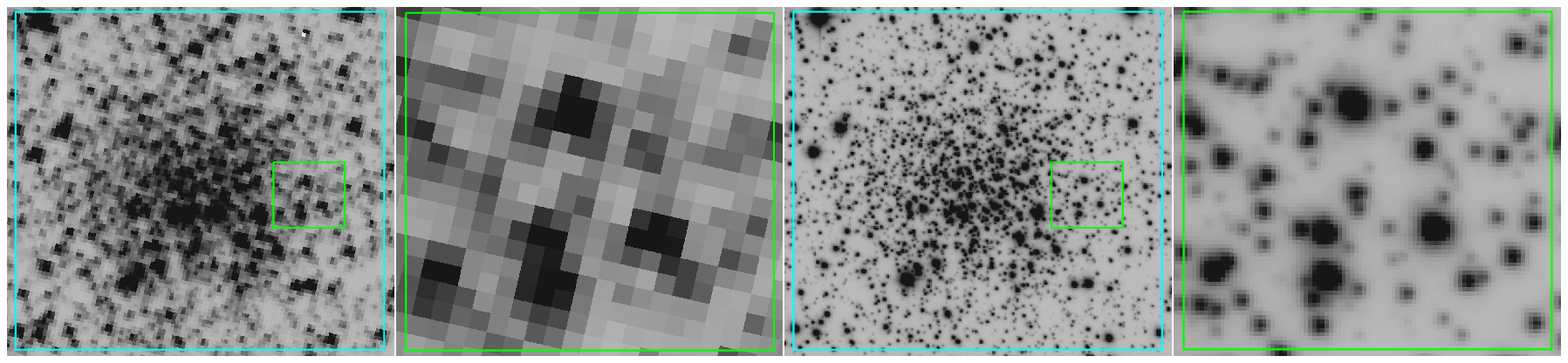}
  \caption{Field of view covered by all available
    \textit{K2}/C0/channel-81 exposures used in our light-curve
    extraction (Sect.~\ref{LC}). On the \textit{Top} we show the
    \textit{K2} stacked image obtained by combining 2422
    exposures. The red rectangle represents the field of view of the
    Schmidt input list of \citet{Nar15}. The cyan and green boxes
    highlight two different zoomed-in regions around NGC~2158. In the
    \textit{Bottom} row we compare these two regions in the
    \textit{K2} (\textit{Left} and \textit{Middle-left} panels) and in
    the Schmidt stacked image from \citet[\textit{Middle-right} and
      \textit{Right} panels]{Nar15}. All these images are oriented
    with North up and East left. It is clear that the higher
    resolution of the Schmidt image allows us to measure more sources
    than in the \textit{K2} image, as well as to better discern
    close-by stars.}
  \label{fig1}
\end{figure*}

\section{Introduction}
\label{intro}

The \textit{Kepler} observatory began science operation in May 12,
2009, with the main scientific goal of discovering exoplanet
candidates transiting their host stars. The \textit{Kepler} Input
Catalog (KIC, \citealt{Brow11}) includes $\sim$150\,000 targets. The
\textit{Kepler} telescope is a defocused 0.95-/1.4-m Schmidt camera
with a field of view of about 100 square degrees.

For a detailed description of the \textit{Kepler} mission and design,
see the \textit{Kepler} technical documents web
page\footnote{\href{https://archive.stsci.edu/kepler/documents.html}{https://archive.stsci.edu/kepler/documents.html}}
and \citet[and references therein]{Koch10}. Here we provide a brief
review of the relevant characteristics. On the telescope focal plane
there is an array of 21 modules, each of which associated to two
2k$\times$1k CCDs. Each CCD is read out in two 1k$\times$1k channels,
for a total of 84 independent channels over the whole focal plane.  On
January 12, 2010, one module failed (MOD-3) and a second one (MOD-7)
failed on February 2014. So, currently only 76 channels are operative.

The pixel scale is $\sim$4 arcsec pixel$^{-1}$, and the deliberate
defocusing causes the average point-spread function (PSF) to extend
across several pixels (except in the area close to the center of the
field). Even so, many features of the PSF are undersampled, even in
the outskirts of the field. Knowing the exact shape of the PSF is the
key for high precision photometry in crowded fields.  The
determination of an appropriate PSF for the \textit{Kepler} images
represents the main effort of this work.

Due to limitation in telemetry, the scientific data were downloaded
once a month using a maximum data-transfer rate of approximately 550
kB/s. For this reason the \textit{Kepler} spacecraft conducts its own
pre-reduction of the data, and sends only a small portion of the
exposed pixels. Around each target, a small area (a ``stamp'') of
various dimensions (typically a few pixels square) is read out with an
integration time of 6.02 s. Every 270 exposures, the stamps are
co-added on board to create a long-cadence stamp of about 29 minutes
of total integration time; short cadences involve adding 9 exposures
for a one-minute integration time.  A time series of such long- or
short-cadence stamps is called a target-pixel file (TPF). When several
targets of interest are in a relatively small patch of sky, multiple
contiguous stamps are collected together to form a so-called
super-stamp. In the original \textit{Kepler} field, two such
super-stamps were taken around the stellar clusters NGC~6791 and
NGC~6819.

In 2012 and 2013, the failures of two reaction wheels that were used
to maintain accurate, stable spacecraft pointing, forced NASA to
re-design the mission. After a period of study and evaluation, NASA
approved an extension to the mission (named \textit{Kepler}-2.0,
hereafter, \textit{K2}, \citealt{How14}), essentially for as long as
the two remaining reaction wheels continue to operate or until the
fuel is exhausted.

The mission was cleverly designed to use the radiation pressure from
the Sun to balance the spacecraft drift, allowing it to observe four
fields per year close to the Ecliptic.  Each of these fields
corresponds to a so called \textit{K2}
Campaign\footnote{\href{http://keplerscience.\-arc.\-nasa.gov\-/K2\-/Fields\-.shtml}{http://keplerscience.\-arc.\-nasa.gov\-/K2\-/Fields\-.shtml}},
and can be continuously observed for $\sim$75 days.

While the two functional reaction wheels control the pitch and yaw,
the thruster needs to be fired every $\sim$6 hours to control the roll
angle of the field. This operation mode causes significantly larger
jitter than in the original \textit{Kepler} main mission. Therefore,
although the \textit{K2} data collection procedures are similar to
those adopted for the original \textit{Kepler} mission, the reduced
pointing capabilities impose the adoption of 2-4 times larger target
stamps.  Because the \textit{K2} stamps include 4-16 times more
pixels, the number of observed targets is proportionally reduced from
the $\sim$150\,000 targets of the original \textit{Kepler} field to
$\sim$10\,000-20\,000 objects for the various \textit{K2} fields. A
new list of target objects is defined for each campaign: the Ecliptic
Plane Input
Catalog\footnote{\href{https://archive.stsci.edu/k2/epic/search.php}{https://archive.stsci.edu/k2/epic/search.php}}
(hereafter, EPIC).

Despite the changes introduced in \textit{K2} mission, several notable
results have been achieved, from exoplanet discovery and
characterization (e.g., \citealt{Cros15,FM15,Van15}) to
asteroseismology (\citealt{Her15,Ste15}) and stellar astrophysics
(\citealt{Arm15,Bal15,Kra15,LaC15}).

\subsection{The purpose of this study}

In this study we apply our expertise on PSF photometry and astrometry
on dithered and undersampled images in crowded fields
\citep{AK00,AK06} to extract high-precision photometry of stars in the
two Galactic open clusters (OCs) M\,35 and NGC~2158, which happened to
lie within the \textit{K2} Campaign 0 (C0) field.

The temporal sampling and coverage of \textit{Kepler} and \textit{K2}
missions and their high photometric precision could be an invaluable
resource for different stellar cluster fields, for example
gyrochronology (\citealt{Bar07,Mam08,Mei11,Mei13,Mei15}), stellar
structure and evolution (e.g., age and Helium content from detached
eclipsing binaries as done by \citealt{Bro11}) or asteroseismology
membership (\citealt{Ste11}). Another interesting topic concerns
exoplanets. We could improve our knowledge about the exoplanets in
star clusters, in particular on how the environment (chemical
composition, stellar density, dynamical interaction) can affect their
formation, evolution and frequency
(\citealt{Gil00,Moc04,Moc06,Adam06,Wel08,Nasc12,Qui12,Qui14,Mei13,Bru14}).

Until now, most of the published studies based on \textit{Kepler} and
\textit{K2} data have focused on isolated, bright objects. Focusing on
\textit{K2} data, photometry on such bright objects is well described
in the literature. The main difference between the methods concern the
mask determination, the stellar centroid measurement and the
subsequent detrending algorithms to improve the photometric precision
(for a review of the \textit{K2} methods adopted by the different
teams, see \citealt{Huang15}). However, the potential scientific
information on faint objects and on stars in the super-stamp crowded
regions has not been completely exploited.

In this paper we intend to obtain the most accurate possible models
for the \textit{Kepler} PSFs and to use them to explore the light
curves (LCs) of the sources in the densest regions that have been and
will be imaged by \textit{Kepler} and \textit{K2}. The \textit{Kepler}
main mission includes 4 OCs (NGC~6791, NGC~6811, NGC~6819, NGC~6866);
many more clusters have been and will be observed during the
\textit{K2} campaigns, which will also include the Galactic center and
the Bulge (Campaign 7 and 9, respectively).

Thanks to our PSF models, for the first time on \textit{Kepler}
images, it will be possible to obtain precise photometry for stars in
crowded fields, and down to $K_{\rm P}$$\sim$24. Having access to
accurate \textit{Kepler} PSFs and comprehensive catalogues from
high-angular-resolution ground-based imaging allows us to subtract
neighbours before measuring the flux of target stars, thus giving us
better corrections for dilution effects that might result in
under-estimates of the true radius of transiting/eclipsing objects.
Indeed, the combination of PSFs and catalogues allow {\it all}
reasonably bright sources within the stamps to be measured accurately.
Finally, even without ground-based imaging, accurate PSFs, combined
with aperture photometry, will allow better identification of blends.
We will illustrate a few examples of mismatched variables in the EPIC
catalogue, showing that the real variable is a different, much fainter
source (see Sect.~\ref{rmslit}).

\section{Image reconstruction}
\label{datared}

This pilot paper makes exclusively use of pixels within channel 81 and
collected during \textit{K2} Campaign 0, focusing on the OCs M\,35 and
NGC~2158. We downloaded all the \textit{K2} TPFs, which contain the
complete time series data, for both these clusters from the ``Mikulski
Archive for Space Telescopes'' (MAST).
In our approach we find it more convenient to work with reconstructed
images of the entire channel, putting all saved pixels into a common
reference frame, rather than working with separate stamps. This gives
us a better sense of the astronomical scene and enables us to work
with all pixels collected at a same epoch in a more intuitive way. We
assigned a flag value to all pixels in the CCD unsaved in any stamp.

We wrote a specialized \texttt{\small PYTHON} routine to construct an
image for each cadence number of the TPFs and the corresponding
\textit{Kepler} Barycentric Julian Day (KBJD) was defined as the
average KBJD of all the TPFs with the same cadence number. For C0
channel 81 we thus reconstructed a total of 2422 usable images. Each
channel-81 image is a 1132$\times$1070 pixel$^2$ array.  The value
assigned to each pixel in each image is given by the \texttt{FLUX}
column of the corresponding TPF. To cross-check that the reconstructed
images were correct, we compared them with full-frame images of the
field (which were also available from the MAST).

Figure~\ref{fig1} gives an overview of our data set. We show the
coverage and the resolution the \textit{K2} stacked image obtained
with 2422 images and the Schmidt stacked image from \citet[see
  Sect.~\ref{AIC} for the description of the catalogue and its
  usage]{Nar15}.

\section{Point-spread function modeling}
\label{PSF}

Even taking into account the large defocusing of the \textit{Kepler}
camera, its PSFs are still undersampled.  Indeed, \textit{Kepler} PSFs
are not simple 2-dimensional Gaussians, and several of the PSF's
fine-structure features are severely undersampled. If not correctly
modeled, these substructures can introduce sources of systematic
errors into the measurement of the positions and fluxes of the studied
sources.  Furthermore, if a PSF model is not sufficiently accurate,
any attempted neighbour-subtraction results in spurious residuals and
consequent systematic errors.

\cite{AK00}, hereafter AK00, introduce a convenient formalism to model
the PSF. Rather than model the \textit{instrumental} PSF as it impacts
the pixels, they model the \textit{effective} pixel-convolved PSF
(ePSF) in a purely empirical way using a simple look-up table
formalism. Their PSF is similar to the \textit{pixel response
  function} (PRF) described in \cite{Bry10}. One of the issues that
AK00 found in modeling the undersampled \textit{Hubble Space
  Telescope} (\textit{HST}) WFPC2 PSF is that such PSFs suffer from a
degeneracy between the positions and the fluxes of the point
sources. An appropriate calibration data set where stars are observed
in different pixels of the detector and maps different sub-regions of
the pixels is required to solve for this issue.

The AK00 approach involves taking a dithered set of exposures and
empirically constructing a self-consistent set of star positions and a
well-sampled PSF model that describes the distribution of light in the
pixels. Such a data set was taken for \textit{Kepler} during
\textit{Kepler}'s early commissioning phase (see \citealt{Bry10} for a
description), but unfortunately it has not yet been made available to
the public (though it may be within a few months according to Thomas
Barclay, private communication).

Given the urgent need for PSFs to make optimal use of the \textit{K2}
data, we decided to do the next best thing to construct an accurate
set of star positions, so that properly sampled PSFs can be extracted
from \textit{Kepler} images. The main-mission \textit{Kepler} data are
not suitable for this, since they have very little dither and each
star provides only one sampling of the PSF. The \textit{K2} data are
actually better for PSF reconstruction purposes, since the loss of the
reaction wheels means that the spacecraft is no longer able to keep a
stable pointing.  Every $\sim$6.5 h, a thruster jet is fired to
re-center the boresight position to its nominal position.  As a
consequence, the stellar positions continuously change during
\textit{K2} observations, with each star sampling a variety of
locations within its central few pixels. Moreover, channel 81 contains
the large super-stamp covering M\,35 and NGC~2158 (our main targets)
with a large number of high signal-to-noise (SNR) point sources.  We
will see in the next sections that this mapping is not optimal, but it
is the best available so far.

\begin{figure*}
  \centering
  \includegraphics[width=\textwidth]{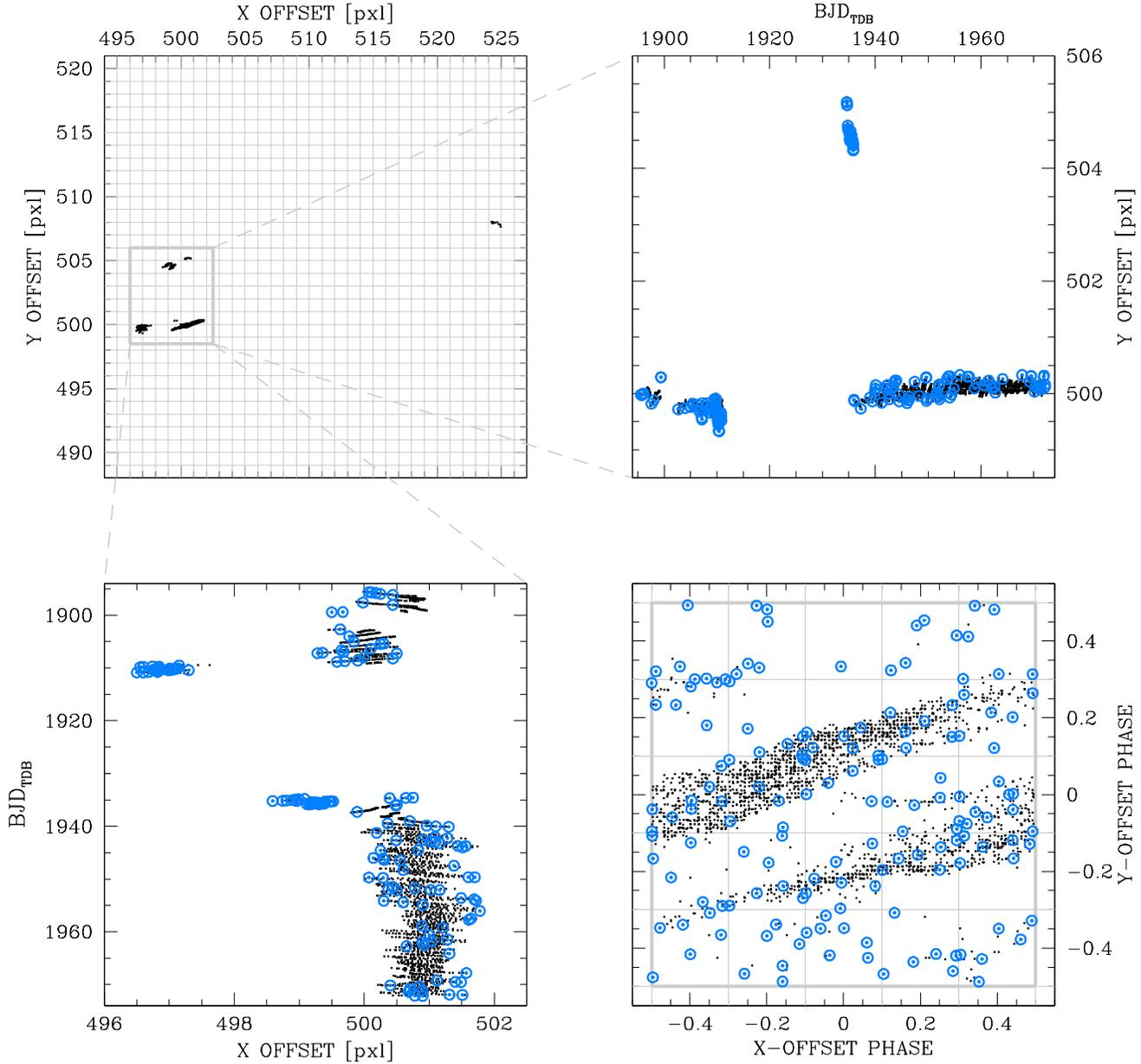}
  \caption{(\textit{Top-left}): dither-pattern outline of the 2422
    images used in our analysis. The gray grid represents the pixel
    matrix in the master-frame reference system. The thick-line gray
    rectangle encloses the points plotted in the \textit{Top-right}
    and \textit{Bottom-left} panels. (\textit{Top-right}): $y$-offset
    variation during C0. We excluded the 10 points around (524,508)
    with the largest offset with respect to the average value to
    better show the time variation of the $y$-offset. The azure points
    are those images used for the ePSF modeling (see
    text). (\textit{Bottom-left}): as on \textit{Top-right} but for
    the $x$ offsets. (\textit{Bottom-right}): dither-pattern pixel
    phase. The center of the pixel (dark gray square) is located at
    (0,0). The pixel was divided into a 5$\times$5-grid (thin gray
    lines) elements and, in each such element, we selected six images
    (when possible) to map that sub-pixel region as homogeneously as
    possible.}
  \label{fig2}
\end{figure*}

\begin{figure*}
  \centering
  \includegraphics[width=\textwidth]{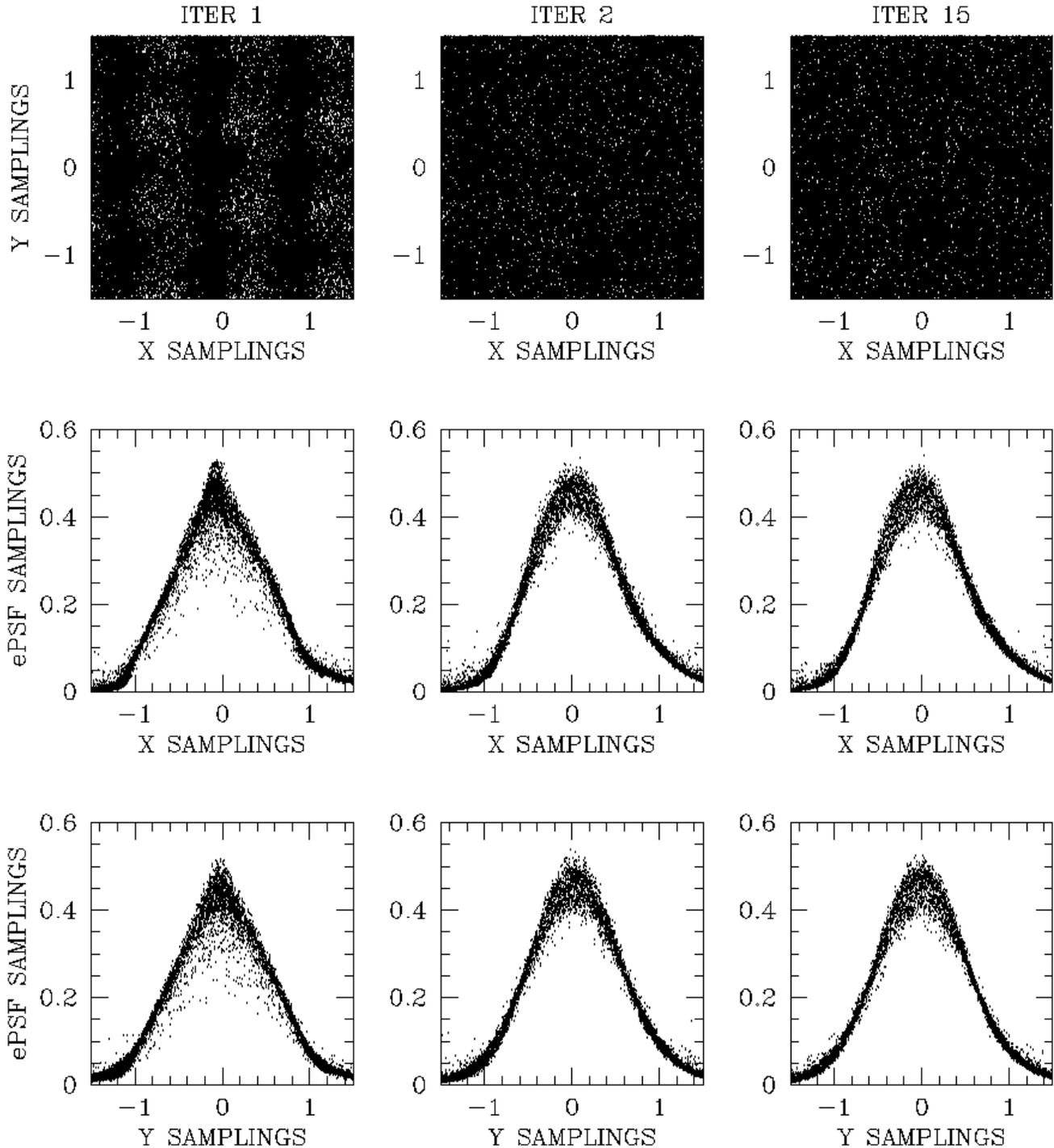}
  \caption{Effective-PSF samplings at the first (\textit{Left}
    column), second (\textit{Middle} column) and last (\textit{Right}
    column) iteration. On the \textit{Top}-row panels we show the
    location of the estimated value of the PSF with respect to the
    center of the stars, placed at (0,0). At the beginning the star
    position was computed using the photocenter. From the second
    iteration, the sampling becomes more uniform. On the
    \textit{Middle}- and \textit{Bottom}-row panels we show the ePSF
    profile along the $x$ and $y$ axes for a thin slice with
    $\mid\Delta y\mid$$<$0.01 and $\mid\Delta x\mid$$<$0.01,
    respectively. In all panels we plotted only 10\% of the available
    points, for clarity.}
  \label{fig3}
\end{figure*}

\begin{figure*}
  \centering
  \includegraphics[width=\textwidth]{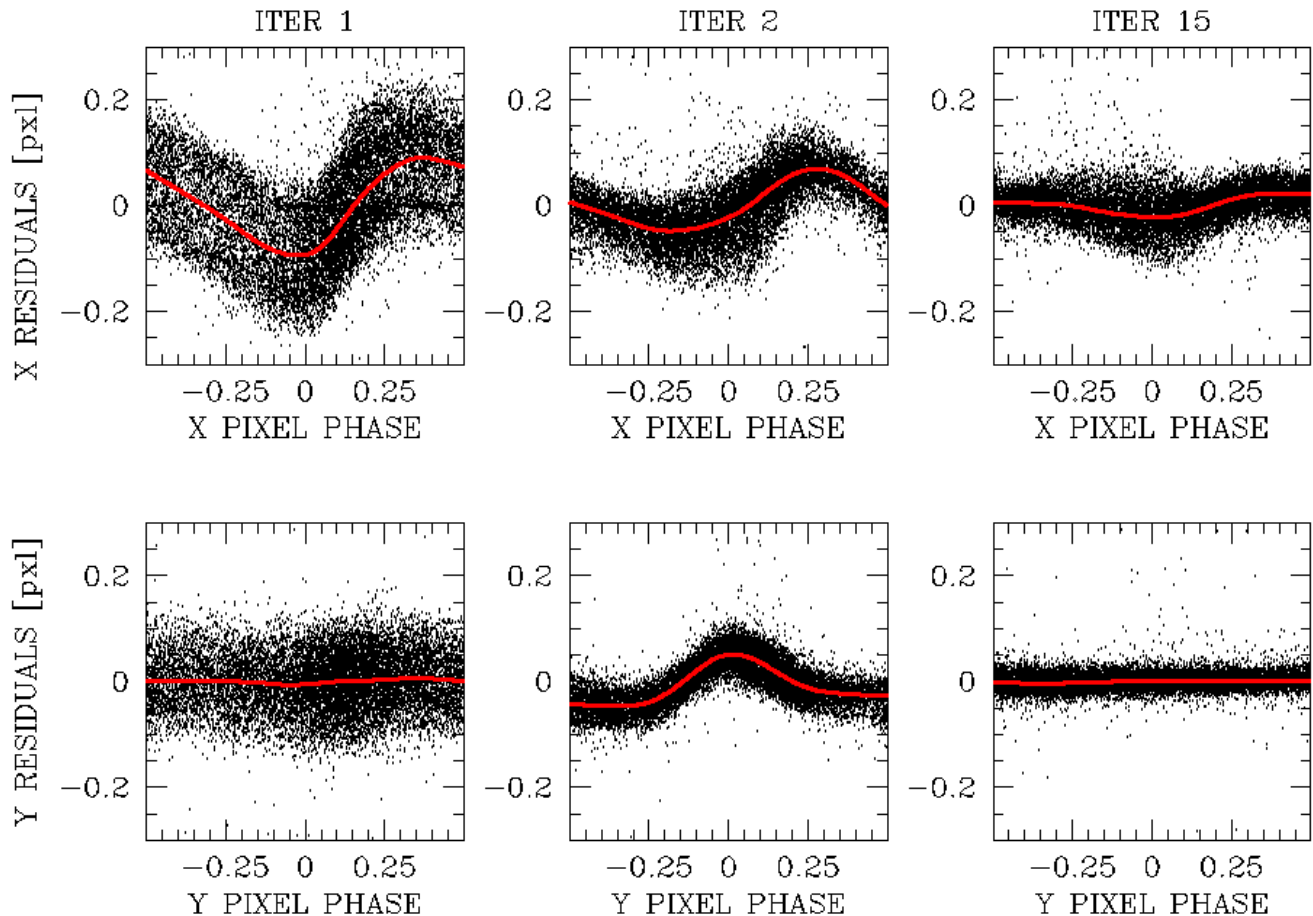}
  \caption{Pixel-phase errors progression. On the \textit{Left} column
    we show the first, in the \textit{Middle} column the second and in
    the \textit{Right} column the last iteration of the ePSF
    modeling. In the \textit{Top} row we show the pixel-phase errors
    along the $x$ axes, in the \textit{Bottom} row the errors along
    the $y$ axes. As in Fig.~\ref{fig3}, we plotted only 10\% of the
    points.}
  \label{fig4}
\end{figure*}

\subsection{Initial assess of the dithered pointings}

The complicated drift-and-repositioning process inherent in the
\textit{K2} data collection results in a very uneven sampling of each
star in pixel phase\footnote{The pixel phase of a star is its location
  with respect to the pixel boundaries and can be expressed as the
  fractional part of its position: ${\rm PIXEL\, PHASE}=x_i-{\rm
    int}(x_i+0.5)$ .}. There are many observations at the initial
phase, and few at the latter phases. We note that even with the
repeated drift, a star typically samples its pixels along a line,
rather than evenly across the face of a pixel, so even the achieved
dither is less than ideal. In order to make the dither sampling as
even as possible, we selected a subset of images (out of the 2422
exposures described in the previous section) in order to evenly map
the astronomical scene across pixel phases. To do this, we used the
\textit{empirical} approach of \cite{Ande06} to construct an initial
PSF model that was spatially constant across each detector. Such a PSF
is not ideal for our ultimate purposes, but it provides better
positions than a crude centroid approach and will allow us to identify
a subset of images that can be used to extract the best possible PSF.

For each exposure, we made one empirical-PSF model for the entire
channel because most of the stars are located in the M\,35/NGC~2158
super-stamp and the spatial variability is not significant for this
initial purpose. With such PSFs, we were able to measure positions and
fluxes for all sources. We then built a common reference system
(hereafter master frame) by cross-identifying the stars in each
image. We used six-parameter linear transformations to bring the
stellar positions as measured in each exposure into the reference
system of the master frame. At the first iteration, the master frame
was chosen using as reference one of the exposures of our sample. We
then adopted a clipped-average of the single-frame positions
transformed into the master frame in order to improve the definition
of the master frame itself, and re-derived improved
transformations. This process was iterated until the master-frame
positions did not significantly change from one iteration to the next.

The transformations between each frame and the master frame allowed us
to analyze the dither-pattern.
In Fig.~\ref{fig2} we show this pattern along with its time
evolution. The dither pattern was made by transforming the same pixel
in each exposure into the master-frame reference system. Since the
behaviour of the geometric distortion is different along the detector,
the dither-pattern outline can change using a different
pixel. However, for our purpose, such representation allows us to
understand it anyway.
It is clear that the dithering places the same star at a range of
locations in its central pixels. While the dither coverage along the
$x$ axis is reasonably uniform, along the $y$ axis the bulk of the
2422 exposures is located in a narrow area. We constructed a
homogeneous mapping of the pixel-phase space (bottom-right panel of
Fig.~\ref{fig2}) as follows. We divided the pixel-phase space in a
5$\times$5-grid elements and, in each element, we selected by hand six
exposures in order to include that sub-pixel region in our PSF
construction. This was possible in almost all the cells. We ended up
with 154 images out of 2422. This is a compromise between the need to
have an adequate number of samplings for the ePSF, and the necessity
to map the entire pixel space homogeneously, avoiding over-weighting
of any sub-pixel region (by using the whole data set which is very
heterogeneous).

\subsection{Building the \textit{effective}-PSF}

With this subset of images, we were finally able to construct a
reliable PSF model. For all exposures, we assumed the PSF to be
spatially constant within our super-stamps, and to have no temporal
variations. For a given star of total flux $z_{\ast}$ located at the
position ($x_{\ast}$,$y_{\ast}$), the value of a given pixel $P_{i,j}$
close to such star is defined as: \looseness=-2
\begin{displaymath}
  P_{i,j}=z_{\ast}\cdot \psi(i-x_{\ast},j-y_{\ast})+s_{\ast}
  \phantom{1} ,
\end{displaymath}
where $s_{\ast}$ is the local sky background and $\psi(\Delta x,\Delta
y)$ is the PSF, defined as the fraction of light that would be
recorded by a pixel offset by $(\Delta x,\Delta
y)=(i-x_{\ast},j-y_{\ast})$ from the center of the star. By fitting
the PSF to an array of pixels for each star in each exposure, we can
estimate its $x_{\ast}$, $y_{\ast}$ and $z_{\ast}$ for each
observation of each star. The equation to be used to solve for the PSF
can be obtained by inverting the above equation: \looseness=-2
\begin{displaymath}
  \psi(\Delta x,\Delta y)=\frac{P_{i,j}-s_{\ast}}{z_{\ast}} \phantom{1} .
\end{displaymath}
With knowledge of the flux and position of each star in each exposure,
each pixel in its vicinity constitutes a sampling of the PSF at one
point in its two-dimensional domain. The many images of stars, each
centered at a different location within its central pixel, give us a
great number of point-samplings and enable us to construct a reliable
ePSF model. This is described in detail in AK00.  Here, we just
provide a brief overview of the key points: \\
$\bullet$ We made a common master frame by cross-identifying bright,
isolated\footnote{The adjective ``isolated'' should be considered in a
  relative way. Within a 1$\times$1-pixel square on a \textit{K2}
  exposure there could be more than one star, as we will see in the
  Fig.~\ref{fig20}. Therefore by ``isolated'' we mean that, in the
  \textit{K2} image, there are not other obvious stars close to the
  target.}  stars in each image and computing their clipped-averaged
positions and flux. On average, we have 650 good stars per exposure to
use. \\
$\bullet$ We transformed these average master-frame positions back
into the frame of each image in order to place the samplings more
accurately within the PSF domain (since each measure is an average of
154 images). \\
$\bullet$ We converted each pixel value in the vicinity of a given
star in a given image into an ePSF sampling, and modeled the variation
of the PSF across its 2-dimensional domain. We have on average 650
reliable stars per exposure, for a total of 10$^5$ samplings. \\
$\bullet$ The available ePSF was used to measure an improved position
and flux for the stars in each image. \\
The whole procedure was iterated fifteen times, after which we noticed
that the overall improvements were negligible (i.e., the pixel-phase
errors did not change from one iteration to the next). The final ePSF
was a 21$\times$21 array of points that maps the 5$\times$5-pixel
region around a star (as in AK00, our ePSF model was supersampled by a
factor 4).

\begin{figure*}
  \centering
  \includegraphics[width=\textwidth]{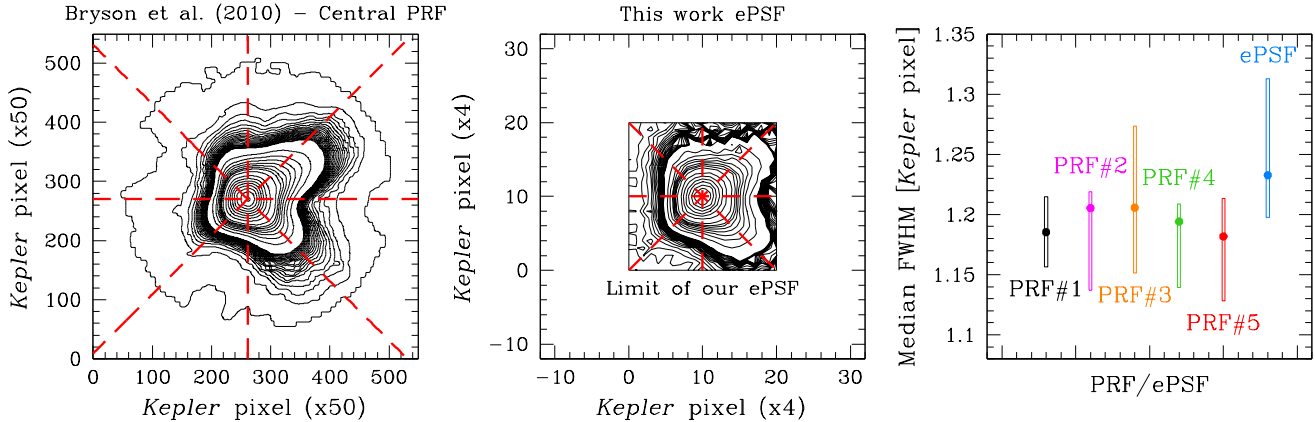}
  \caption{FWHM analysis. In the \textit{Left} and \textit{Middle}
    panels, we show a contour representation of channel-81 central PRF
    and our average ePSF, respectively. The red dashed lines show the
    direction along which we measured the FWHM ($-$45, 0, 45, and 90
    degrees). The PRF and the ePSF are plot with their original
    supersampling factor (50 and 4, respectively) but their size in
    the images are the same (in real \textit{Kepler} pixels). It is
    clear that our ePSF is limited to the center part, due to the data
    set used that was not designed to model it. In the \textit{Right}
    panel, we show in different colours the median of these FWHM
    values for the five PRFs (black, magenta, orange, green and red)
    and for our \textit{K2}/C0 ePSF (azure). The rectangle around each
    point shows the minimum-to-maximum range of the FWHM values along
    the considered directions.}
  \label{figpsfcomp}
\end{figure*}

In Fig.~\ref{fig3} and \ref{fig4}, we show the result of our
procedure. At the beginning, the ePSF sampling is not homogeneous and
the shape of the ePSF is not well constrained. Even in the second
iteration, the sampling began to improve and the ePSF became
smoother. The same behavior can be seen in the pixel-phase
errors. Note that the pixel-phase errors appear to be behaved along
the $y$ axes rather than the $x$ axes. However, it is not clear
whether the available pixel-phase sampling simply allows us to see the
errors in $x$ better than those in $y$. Again, when the
\textit{Kepler} PSF-characterization data set becomes public, it will
allow for a much better characterization and verification of the PSF.

\subsubsection{Comparison with \textit{Kepler}-main-mission PRFs}

Following a suggestion by the referee, we investigated whether our
\textit{K2}/C0 ePSF is broader than the \textit{Kepler}-main-mission
PRF (\citealt{Bry10}). The broadening is expected because of the
\textit{K2} pointing jitters larger than those in the main mission.

We measured the full-width half-maximum (FWHM) of the five channel-81
PRFs (one for each corner and one for the central region of the CCD)
and that of our average ePSF along different directions.  In
Fig.~\ref{figpsfcomp}, we show a contour representation of our ePSF
and the central PRF of \cite{Bry10}. The ePSF size in these
representation is the same as for the PRF (in \textit{Kepler}
pixels). Our ePSF is limited to a small region around the center and
do not model the ePSF wings, as the crowding in the studied
super-stamp does not allow the modeling of the ePSF wings. In the
right panel of Fig.~\ref{figpsfcomp} we show the median FWHM values
for the PRFs/ePSF. We found that our \textit{K2}/C0 ePSF is broader
than the \textit{Kepler}-main-mission PRFs.

\subsection{Perturbed ePSF}

The basic assumption of the original AK00 method is that the ePSF is
constant in time and identical for all the images.  This is of course
an ideal condition and ---at some level--- it is never the case;
surely not for the \textit{HST} nor for any other instrument we have
used.  There are also other subtle effects.  The selection of a
uniformly-distributed subsample of images in pixel-phase space could
have introduced some biases.  For example, some of the dither
pointings in the less-populated regions of the pixel-phase space could
have been taken while the telescope was still in motion, resulting in
a more ``trailed'' ePSF than the average ePSF. In any case, as a
working hypothesis, and having detected no obvious trailed images in
the subsample that we selected for the ePSF determination, we
proceeded under the approximation that these effects are not larger
than the general, ePSF variations as a function of time. Indeed, the
shape of the \textit{Kepler} ePSF clearly changes over time, as one
can infer in Fig.~\ref{fig3} from the vertical broadening of the ePSF
around the center (rightmost middle and bottom panels).

\begin{figure}
  \centering
  \includegraphics[width=\columnwidth]{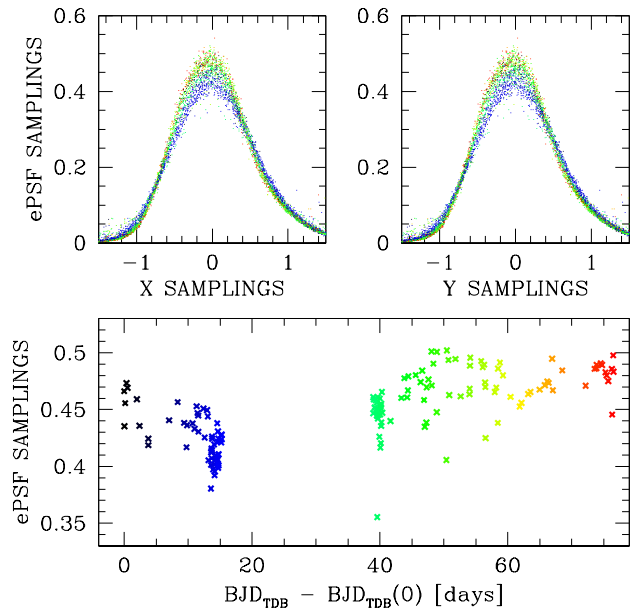}
  \caption{Time variation of the ePSF shape in the 154 images we used
    for the modeling. In the \textit{Bottom} panel, the central-peak
    value of the ePSF (interpolated from the samplings) as a function
    of the time interval relative to the first-image epoch. In the
    \textit{Top-left} and \textit{Top-right} panels we show the ePSF
    samplings as in the \textit{Right} column of Fig.~\ref{fig3},
    using the same colour-codes for the time of image acquisition.}
  \label{fig5}
\end{figure}

Figure~\ref{fig5} illustrates the temporal variation of the PSF.  We
colour-coded the samplings of the final ePSF (top panels) according to
the epoch of observation (see bottom panel). There are clear trends
that are not simply monotonic.

In order to suppress as much as possible the impact of the temporal
dependencies of the ePSF we introduced a perturbation of the average
ePSF.  This perturbation of the ePSF was first described in
\cite{AK06}, and can be summarized as follows.  In each image, we fit
and subtracted the current ePSF model to high SNR stars.  We then
modeled (with a look-up table) the normalized residuals of these ePSF
fits and added these tabulated residuals to the initial ePSF to obtain
an improved ePSF model (for a more recent application and a detailed
description of the method see \citealt{Bel14}).

\begin{figure}
  \centering
  \includegraphics[width=\columnwidth]{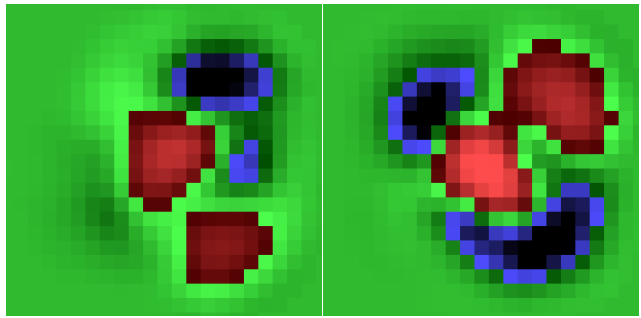}
  \caption{Difference between the average ePSF and the perturbed ePSF
    for an image at the beginning (\textit{Left}) and at the end
    (\textit{Right}) of Campaign 0. In these two examples, the
    variation ranges between $\sim$$-0.5$ per cent (blue colour) and
    $\sim$1.4 per cent (red colour) of the total flux.}
  \label{fig6}
\end{figure}

\begin{figure}
  \centering
  \includegraphics[width=\columnwidth]{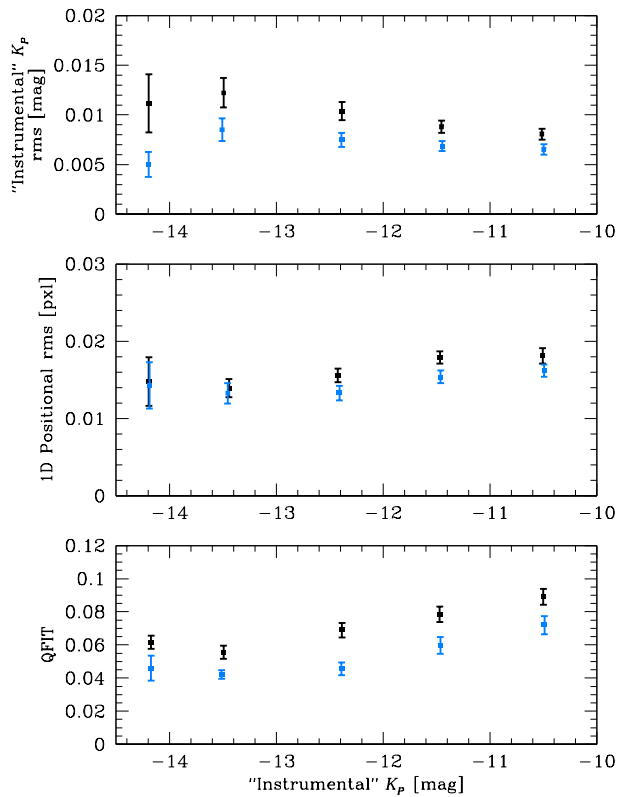}
  \caption{\textsf{QFIT} (\textit{Bottom}), 1D positional rms
    (\textit{Middle}) and photometric rms (\textit{Top}) as a function
    of the ``instrumental'' magnitude of the master frame with (azure
    points) and without (black points) perturbing the average ePSF to
    account for the time-dependent variations. Each point represents
    the median value in 1-magnitude bins. The error bars represents
    the 68.27$^{\rm th}$ percentile of the distribution around the
    median value divided by $\sqrt{N}$, with $N$ number of points. The
    instrumental magnitude in each single catalogue is defined as
    $-2.5\times \log (\sum \rm{counts})$, where $\sum\rm{counts}$ is
    the sum of the total counts under the fitted PSF. Since we used
    the column \texttt{FLUX} while building the full-frame channel
    images, these counts are actually fluxes (for this reason we used
    the double quotes in the label). Hereafter, we will omit the
    double quotes in the text.}
  \label{fig7}
\end{figure}

In Fig.~\ref{fig6} we show an example of the time-variation
adjustments of the average ePSF for two images. The improvements in
position and flux of the perturbed ePSFs over a constant ePSF for all
the 2422 images are quantified in Fig.~\ref{fig7}. We measured
position and flux of all sources in each exposure with and without
perturbing the ePSF. The PSF-fit process (a least-square fit) is
achieved with a program similar to the \texttt{\small img2xym\_WFI}
described in \cite{Ande06} that measure all sources, from the
brightest to the faintest (up to a minimum-detection threshold set by
the user) objects, in seven iterations. From the second iteration
ahead, the fitting procedure is also performed on neighbour-subtracted
images, in order to converge on reliable positions and fluxes for
close-by objects. We then made two master frames, one for each of the
two different approaches, by cross-identifying the stars in each of
our 2422 images. We computed the 3$\sigma$-clipped median value of the
following quantities: \textsf{QFIT}\footnote{The \textsf{QFIT}
  represents the absolute fractional error in the PSF-model fit to the
  star \citep{Ande08}. The lower the \textsf{QFIT}, the better the PSF
  fit.}, the 1D positional rms, and the photometric rms.  These values
were calculated in 1-magnitude bins and, for an appropriate
comparison, we used the same set of stars for the two samples. In most
cases, we found that the difference between the use of perturbed or
unperturbed ePSF is not negligible. Using the perturbed rather than
the average ePSF to measure position and flux of a star in an image
improves the PSF fit because the former is a representation of a star
in that particular image, while the latter is the representation of a
star averaged on the entire C0.

In the following, we assume that the spatial variation of the ePSF
across the super-stamp region in channel-81 detector where NGC~2158
and M\,35 are imaged is negligible, therefore we will use only one
ePSF model per image. A substantial improvement to the ePSF will be
achieved when the PSF characterization data are made available, so
that we can properly account for the spatial variability.

\section{Photometry in \textit{K2} reconstructed images}
\label{LC}

The main purpose of this effort is to extract precise photometry from
main-mission and \textit{K2}-mission pixel data for sources in crowded
fields and at the faintest end. These two issues are very closely
related, since each 4$^{\prime\prime}$$\times$4$^{\prime\prime}$ pixel
includes many faint sources that we need to take into account. Because
of this crowding, classical aperture photometry has major obvious
limitations.

Different authors showed that photometry on neighbour-subtracted
images leads, on average, to a better photometric precision than on
the original images (see, e.g., \citealt{Mon07} and reference
therein). By subtracting a star's neighbours before measuring its
flux, it is possible to avoid including (or at least to reduce the
impact of) light that could contaminate the true flux of the
target. The knowledge of positions and flux of all sources in the
field is therefore fundamental to our approach.

In order to obtain the best photometric precision with \textit{K2}
data, we used the same method as described in \cite{Nar15} to which we
refer for a more detailed description. This method makes use of
accurate PSF models for each exposure and of a (ground-based) input
list to disentangle from the flux of a given star the contribution of
its close-by sources. The more complete the input list with all
detectable objects in the field, the better will be the final result
of the method. A corollary is also that the transformations of the
input-list positions and fluxes into the individual-exposure reference
system need to be known with high accuracy, as well as the PSF, in
order to subtract the neighbours as well as possible. In the following
we provide the adopted key ingredients.

\subsection{The Asiago Input Catalogue (AIC)}
\label{AIC}

Our input list is described in great detail in \cite{Nar15}.  It comes
from a set of white-light (i.e., filterless) observations collected at
the Asiago Schmidt telescope. It includes 75\,935 objects. At variance
with \citeauthor{Nar15}, we used all stars measured in white light,
and not only those found in both white-light and $R$-filter lists (see
Sect.~3.5 of their paper). Hereafter, we will refer to this catalogue
as the \textit{Asiago Input Catalogue}, or AIC, which is available at
the ``Exoplanet and Stellar Population Group'' (ESPG)
website\footnote{\href{http://groups.dfa.unipd.it/ESPG/M35NGC2158.html}{http://groups.dfa.unipd.it/ESPG/M35NGC2158.html}}.

The AIC was constructed by measuring the position and flux of each
source found in the Schmidt stacked image via PSF fitting. This input
list was then transformed into the photometric system of a reference
image (among those of the Schmidt data set). The catalogue was purged
for various artifacts, such as PSF bumps and fake detections along the
bleeding columns. The purging is not perfect and it is a compromise
between excluding true, faint stars around bright objects, and
including artifacts in the catalogue. Of the 75\,935 sources included
in the input list, $\sim$77$\%$ could be measured reasonably well with
the \textit{K2} data set.

The stacked Schmidt image has higher angular resolution than the
\textit{Kepler}/\textit{K2} images, allowing us to locate faint
sources whose flux could pollute the pixels of a nearby star. The
relative astrometric accuracy of the AIC is also sufficiently accurate
to allow us to pinpoint a star in any given \textit{K2} image with an
accuracy down to about 20 mas (0.005 \textit{Kepler} pixel). Details
about the absolute astrometry, the stacked images, and other
information of the AIC can be found in \cite{Nar15}.

\subsection{Photometry with and without neighbours}

The procedure for LC extraction is the same as in \cite{Nar15} and can
be summarized as follows. For each star in the AIC, hereafter ``target
star'', we computed six-parameter, local linear transformations to
transform the AIC position of the target into that of each individual
\textit{K2} image. In order to compute the coefficients of the
transformations, we used only bright, unsaturated, well-measured
target's neighbours within a radius of 100 pixels (target star
excluded). If there were not at least 10 neighbour stars within such
radius, we increased the searching area progressively, out to the
whole field. Local transformations were used to minimize the effect of
the geometric distortion, since it does not vary significantly on
small spatial scales.

These reference stars were also used to transform the AIC white-light
magnitude of the target into the corresponding instrumental
\textit{K2} magnitude in each given exposure. The AIC magnitudes in
white light and our $K_{\rm P}$ instrumental magnitudes are obtained
from instruments with a rather similar total-transmission curve and,
as a first approximation, we can safely use a simple zero-point as
photometric transformation between AIC and \textit{K2} photometric
systems.

We extracted the photometry from the original and from the
neighbour-subtracted \textit{K2} images. The neighbour-subtracted
images were created by subtracting from the original images the
flux-scaled perturbed ePSF of all AIC sources within a radius of 35
\textit{Kepler} pixels ($\sim$2.3 arcmin) from the target star. We
postpone the discussion about the quantitative improvements in the
photometry by using neighbour-subtracted images instead of the
original images to Sect.~\ref{photprecwwo}.

On the neighbour-subtracted images, the target flux of each AIC object
was measured using aperture photometry with four aperture sizes (1-,
3-, 5- and 10-pixel radius) and using ePSF-fit photometry. On the
original images, we used only 3-pixel-aperture and ePSF-fit
photometry. These fluxes were stored in a LC file, which also contains
other information such as the KBJD.

\begin{figure*}
  \centering
  \includegraphics[width=\textwidth]{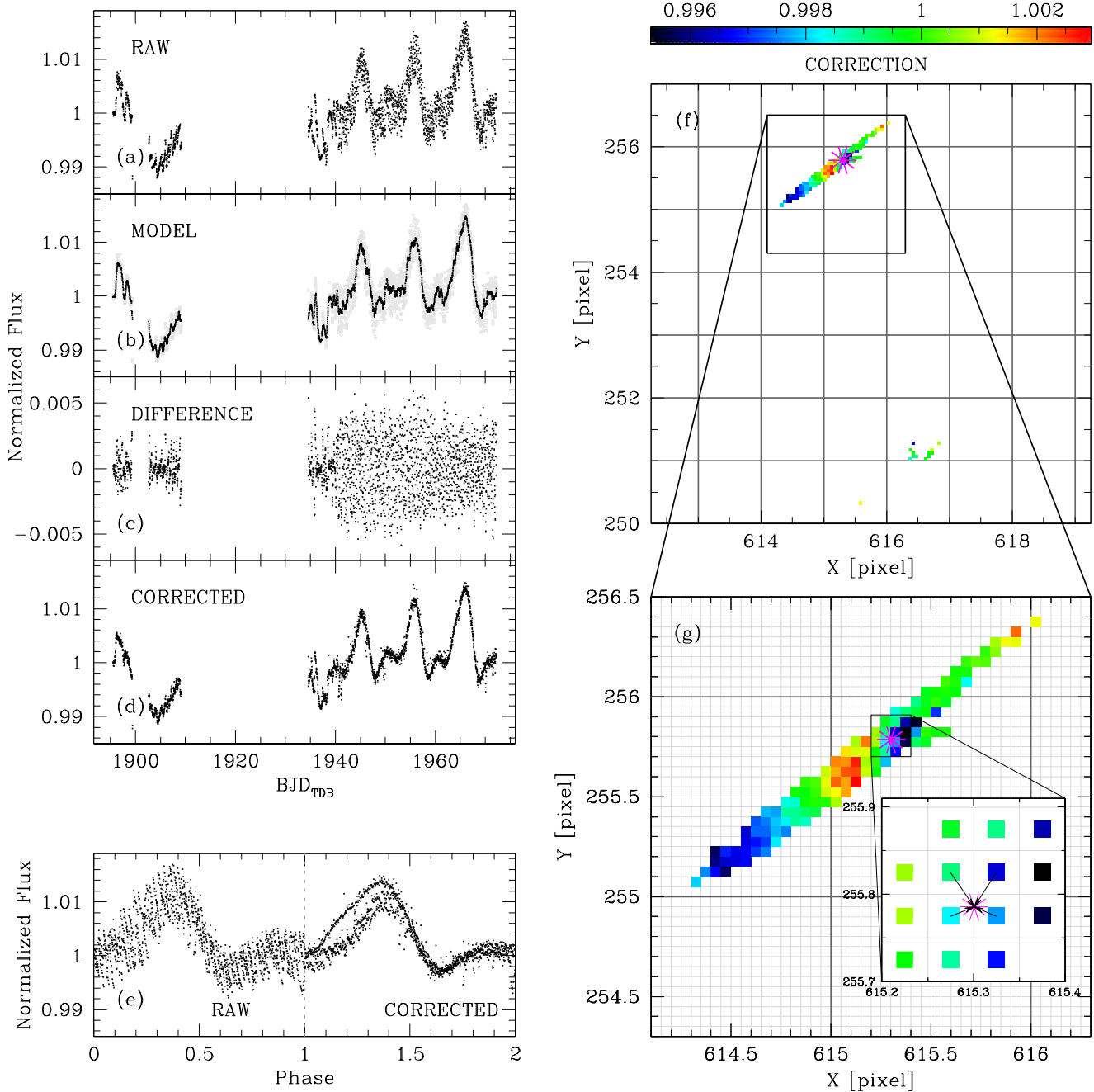}
  \caption{(\textit{Left}): 3-pixel-aperture LC of star \# 1881
    (EPIC~209183478) after each step of our correction. From panel (a)
    to (d): raw LC normalized to its median flux; LC model (black
    points) put on top of the raw LC (light-gray points);
    model-subtracted LC; final detrended LC. In panel (e) we show the
    comparison between the phased (with a period of $\sim$10.6 days)
    LC before (left) and after (right) the correction. It is clear
    that the drift-induced scatter in the LC is reduced and more
    details arise, e.g., the LC amplitude of this object changed
    during the C0. (\textit{Right}): Outline of our correction. In
    panel (f) we show the cell and grid-point locations around the
    star \# 1881 loci on the channel 81 over the entire Campaign
    0. The thick, dark-gray lines mark the pixel boundaries. The
    magenta $*$ marks the star location on the channel at a given
    time. The squares represent the median values of the
    model-subtracted flux in each of the 0.05$\times$0.05-pixel-square
    bins in which each pixel is divided, colour-coded accordingly to
    the colour bar on top. In panel (g) we zoom-in around the bulk of
    the points to highlight the sub-pixel elements of the grid (thin,
    light-gray lines). For the star position at a given time, we used
    the four surrounding grid points to perform the bi-linear
    interpolation (sketched with the arrows) and compute the
    correction.}
  \label{fig8}
\end{figure*}

\section{Photometric calibration into the $K_{\rm P}$ System}
\label{KPphot}

In order to calibrate our the instrumental magnitudes we extracted
from the individual \textit{K2} exposures to the \textit{Kepler}
Magnitude System ($K_{\rm P}$), we determined and applied a simple
local zero-point. For each photometric approach (1-, 3-, 5-, 10-pixel
aperture and PSF), we made a catalogue with absolute positions from
the AIC and with magnitudes obtained from 3$\sigma$-clipped median
values of the \textit{K2} instrumental magnitudes as measured in each
LC (when available). We then cross-matched these catalogues with the
EPIC obtained from the MAST archive. We computed the zero points as
the 3$\sigma$-clipped median value of the difference between our
magnitudes and the EPIC $K_{\rm P}$ magnitudes. We used only those
bright, unsaturated stars that in our catalogue are within three
magnitudes from the saturation and for which the $K_{\rm P}$ magnitude
in the EPIC was obtained from `gri' photometry. We chose this specific
photometric method among the different methods adopted to compute the
$K_{\rm P}$ magnitude in the EPIC due to the larger number of sources
in common between this EPIC subsample and our well-measured-star
sample. As in \cite{Aig15} and \cite{Lund15}, the zero-points of our
photometric methods we found are between 24.7 (1-pixel aperture) and
25.3 (the other photometric methods).

\section{Detrending of \textit{K2} light curves}
\label{detrend}

The unstable spacecraft pointing results in a well-defined motion of
the stars across the pixel grid. A combination of intra- and
inter-pixel sensitivity variation leads to a correlation between this
motion and the ``raw'' flux of each star that must be corrected in
order to increase the \textit{K2} photometric accuracy and
precision. Different methods have been developed to correct such
systematic effect, e.g., the self-flat-fielding approach of
\cite{V&J14}, the Gaussian process of \cite{Aig15} or the simultaneous
fit for all systematics of \cite{FM15}.

In order to detrend the LCs from the drift-induced effects, we took
into account all usable channel-81 exposures collected during the
entirety of C0, including those taken during the first part of the
campaign, when the fine guiding was still in progress (which caused
the stars to be shifted by up to 20 pixels from their average
position, see Fig.~\ref{fig2}). This is important, since the number of
points to be used for the detrending increases, and it could also be
useful to detect variable stars with periods of $\sim$35 days (the
duration of the C0 after the second safe mode). Briefly, our
correction was performed by making a look-up table of corrections and
applying it with simple bi-linear interpolation.

An overview of our detrending approach is shown in
Fig.~\ref{fig8}. For each target star, we made a model of the raw LC
trend, normalized by its median flux (panel a), by applying a
running-average filter with a window of $\sim$10 hours. The model
(panel b) was then subtracted from the raw LC with the aim of removing
the intrinsic stellar variability. In this way, the model-subtracted
LC (panel c) reflects the systematic effect originated from the motion
of the spacecraft. The window size of the running-average filter was
chosen as a compromise between our attempt of avoiding the removal of
the positional trend and still being able to model short-period
variables.

Each pixel into which a given star fell during the motion was divided
into an array of 20$\times$20 square elements and, in each such
element, we computed the 3.5$\sigma$-clipped median of the
model-subtracted LC flux.  The grid is sketched in panel (f) and (g)
of Fig.~\ref{fig8}. For any location on the CCD, the correction was
performed by applying a bi-linear interpolation between the
surrounding four grid points. The correction was not always available,
because for some grid elements there are not enough points to compute
the correction. In these cases, no correction was applied.

The whole procedure was iterated three times, each time making an
improved LC model by using the LC corrected with the available
detrending solution. In Fig.~\ref{fig8} we show the results of the
final iteration of our procedure.

Panel (e) of Fig.~\ref{fig8} shows a direct comparison of the folded
LC before (left) and after (right) the correction. The rms is improved
and allows us to see more details in the LC.

One advantage of our LC-extraction method is a more robust position
measurement. Indeed, as described in Sect.~\ref{LC}, we transformed
the position of a target star as given in the AIC into the
corresponding image reference system by using a subset of close-by
stars, target excluded, to compute the coefficients of the
six-parameter linear transformations. The local-transformation
approach reduces (on average by a factor $N^{-1}$, where $N$ is the
number of stars used) most of the systematic effects that could harm
the stellar positions (e.g., the uncorrected geometric distortion).

\begin{figure*}
  \centering
  \includegraphics[width=\textwidth]{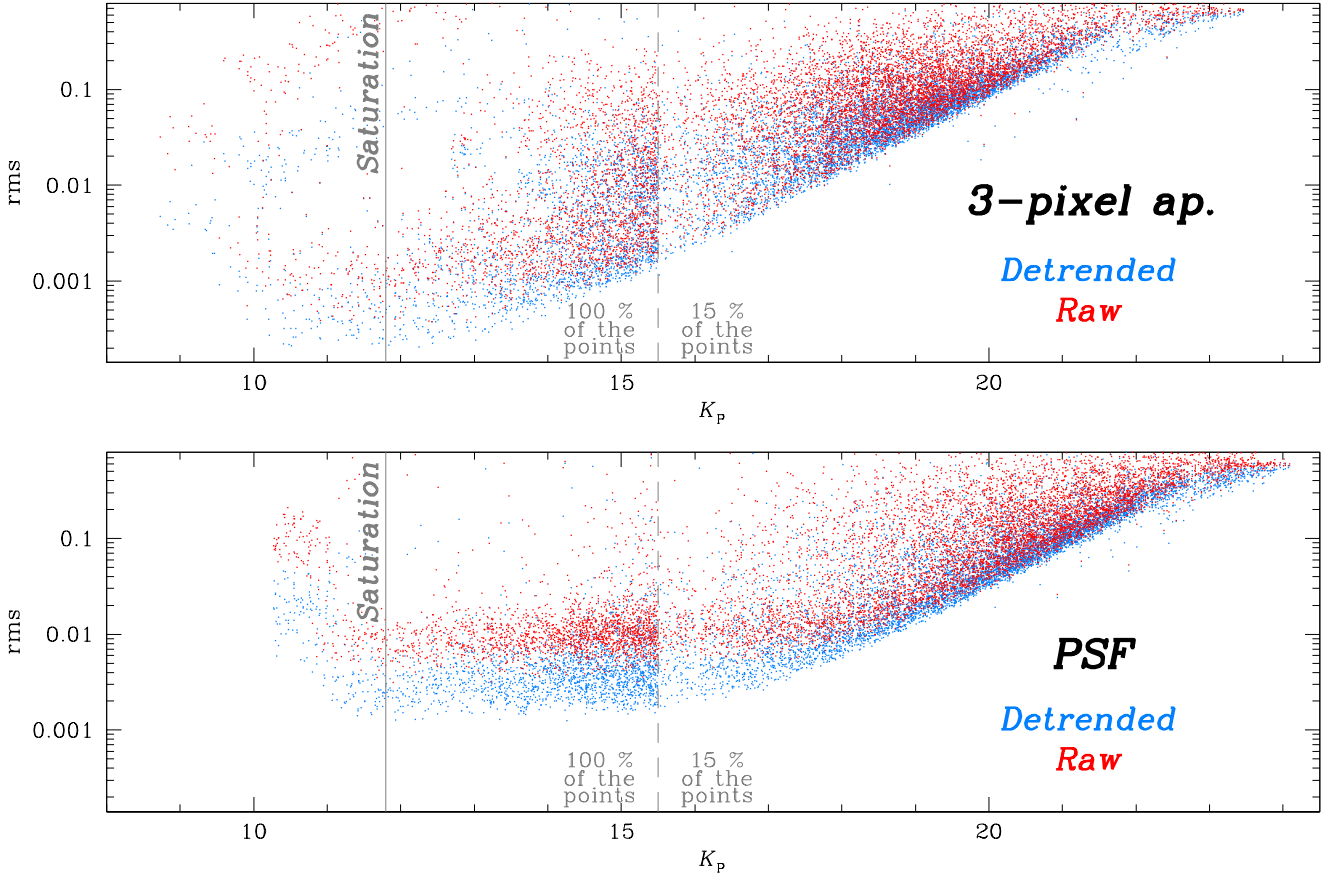}
  \caption{Photometric rms as a function of the $K_{\rm P}$ magnitude
    before (red points) and after (azure points) applying our
    detrending procedure. We plot only the neighbour-subtracted
    photometry from the two measurements that show the best rms in the
    bright (3-pixel aperture, \textit{Top} panel) and faint (PSF,
    \textit{Bottom} panel) regime. For $K_{\rm P}$$>$15.5 (vertical,
    gray dashed line) we show only 15\% of the points for clarity. The
    vertical, gray solid line is set at the saturation threshold
    $K_{\rm P}$$\sim$11.8.}
  \label{fig9}
\end{figure*}

The LC detrending is critical both to the removal of the
spacecraft-related systematics that degrade the \textit{K2}
photometric precision and to pushing \textit{K2} performance as close
as possible to that of the original \textit{Kepler} main mission. In
Fig.~\ref{fig9} we show the rms (defined as the 3.5$\sigma$-clipped
68.27$^{\rm th}$-percentile of the distribution about the median value
of the points in the LC) improvement after the detrending process for
the 3-pixel-aperture and PSF photometry on neighbour-subtracted
images. The rms was calculated on the normalized LC, i.e. after the LC
flux was divided by the 3.5$\sigma$-clipped median value of all the
flux measurements. As we will see in sub-section \ref{photprec3psf},
these two photometric methods are the best-performing methods at the
bright- and faint-end regime, respectively. The improvement is greater
for bright stars, i.e. stars with high SNR and better-constrained
positions; while for faint stars the effect is lost in the random
noise.

\subsection{A previously unknown 2.04-day periodic artifact}
\label{204d}

\begin{figure*}
  \centering
  \includegraphics[width=\textwidth]{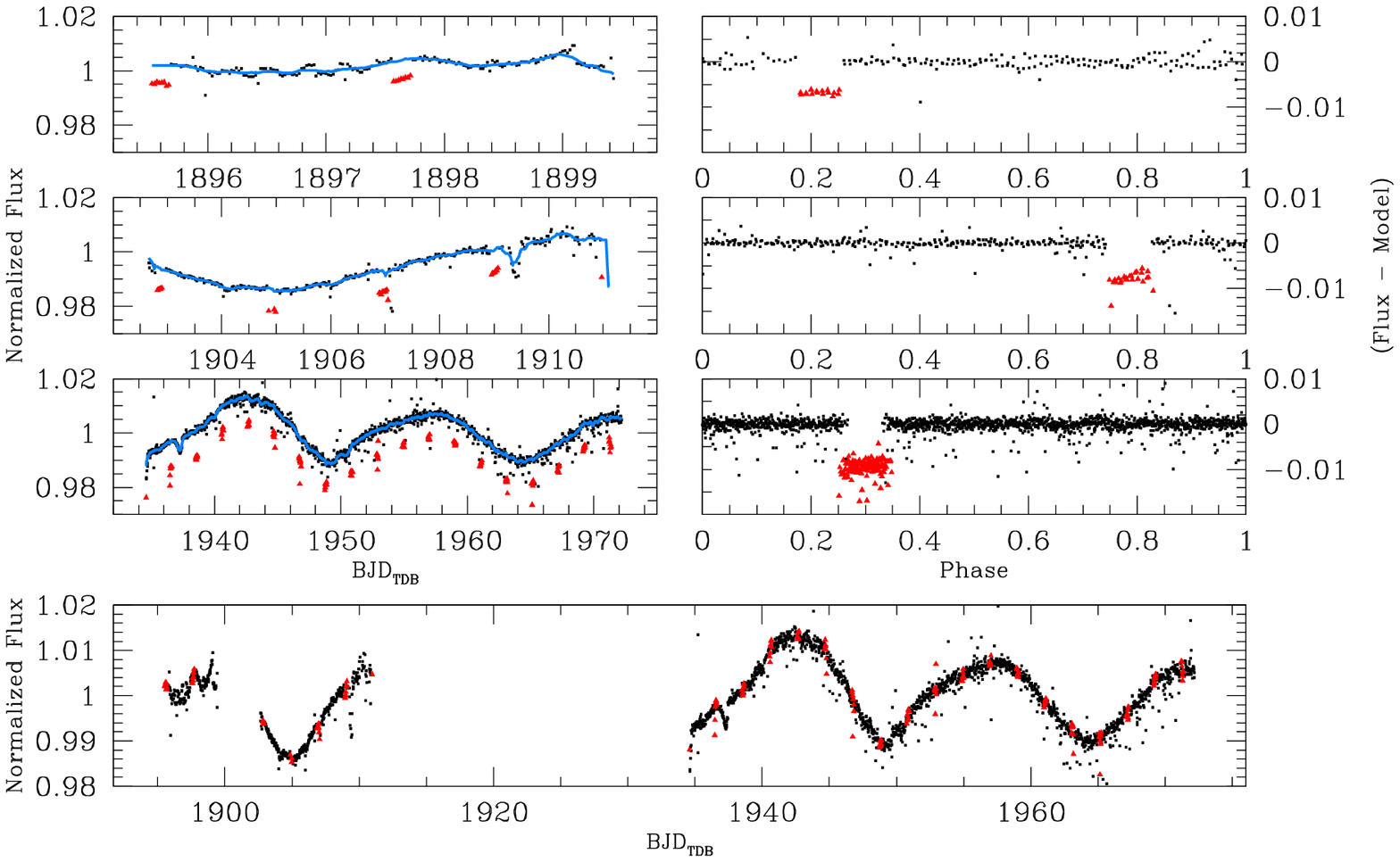}
  \caption{Example of a LC (AIC \# 8600, 5-pixel-aperture photometry)
    affected by the flux-drop effect described in
    Sect.~\ref{204d}. From \textit{Top}, in the first-row panels we
    show the first sub-series (before the first safe mode) of points
    of the C0/channel-81 LC (\textit{Left}) and the corresponding
    phased LC with a period of 2.04 days (\textit{Right}). In the
    second and third rows we show the same as above but for the second
    (between the first and the second safe modes) and third (after the
    second safe mode) sub-series of the LC and the phased LC. The
    black dots represent the unaffected points, the red triangles are
    the photometric points affected by the flux drop. The azure line
    is the LC model. Note that the phase at which the flux drop occurs
    is not the same in the different parts of the LC, meaning that
    this effect could be related to the instrumentation, and that it
    is reset after each break in the C0. Finally, in the
    \textit{Bottom} panel we plot the corrected LC.}
  \label{fig10}
\end{figure*}

Several LCs turned out to exhibit a periodic drop in their fluxes that
we have not seen described anywhere. These drops are rather sharp and
resemble boxy transits; we illustrate them in Fig.~\ref{fig10}. The
period of this effect is $\sim$2.04 days, and the drops last for
exactly six images, for each of the non-interrupted sub-series within
C0. Not all the stars in the affected images show the drop feature.

We extensively investigated the affected and un-affected stars on
individual \textit{K2}/C0/channel-81 images and found that the effect
could be column-related. The amount of drop in the flux is not always
the same and it correlates with magnitude. We have not found any
description of such an effect either in the \textit{Kepler} manual or
in the literature.

We suspect that it might be due to an over-correction in
correspondence of mis-behaving columns. This over-correction might be
the result of electronic activities related to the periodic momentum
dump of the two remaining reaction wheels through thruster firings;
which happens every two days (\citealt{How14,Lund15}). For example, it
could be a change in the reading rate.

Another possibility, discussed with the referee, is that this effect
can be originated by a non-identical \textit{Kepler}-pipeline
processing of contiguous TPFs that can create some un-physical
discontinuity. Such 2.04-d periodic effect could only be detected when
dealing with more than one TPF at a time, as we did in our
reconstructed-image approach.

Lacking more engineering data and detailed knowledge of the on-board
pre-reduction pipeline, we have limited ourselves to simply describing
and correcting these effects empirically. We first made a model of the
LC trend (by applying a running-average filter with a window of
$\sim$6 hours) and computed the median value of the model-subtracted
LC flux for affected and un-affected data points. If the difference
between these median values was greater than the model-subtracted LC
rms (computed using only the un-affected points), we marked the LC as
flux-drop-affected, and corrected it. The correction to add is
computed as the difference between the model of the LC with only the
un-affected points and the model obtained using only the
flux-drop-affected points (see Fig.~\ref{fig10}).

The drawback of our correction is obviously that any true variable
star with a significant flux drop every 2.04 days was considered as
flux-drop affected and corrected accordingly. Of course, since this
flux-drop is not periodic across safe-mode sub-series, it is possible
to distinguish these events from true eclipses.

\subsection{The role of \lowercase{e}PSF\lowercase{s} in \textit{K\lowercase{epler}}/\textit{K2} images}
\label{PSFrole}

The ePSF is not only more suitable to performing photometry in crowded
regions and for faint stars, but it can also be used as an additional
diagnostic tool to discern among exposures that are most affected by
some systematic effects, such as the drift motion of the
spacecraft. In the following we show that, by taking advantage of all
the information that we have from the ePSFs adjusted for each
exposure, we can also select the best exposures (that correspond to
points in the LCs) to search for variability (Sect~\ref{search}).

\begin{figure*}
  \centering
  \includegraphics[width=\textwidth]{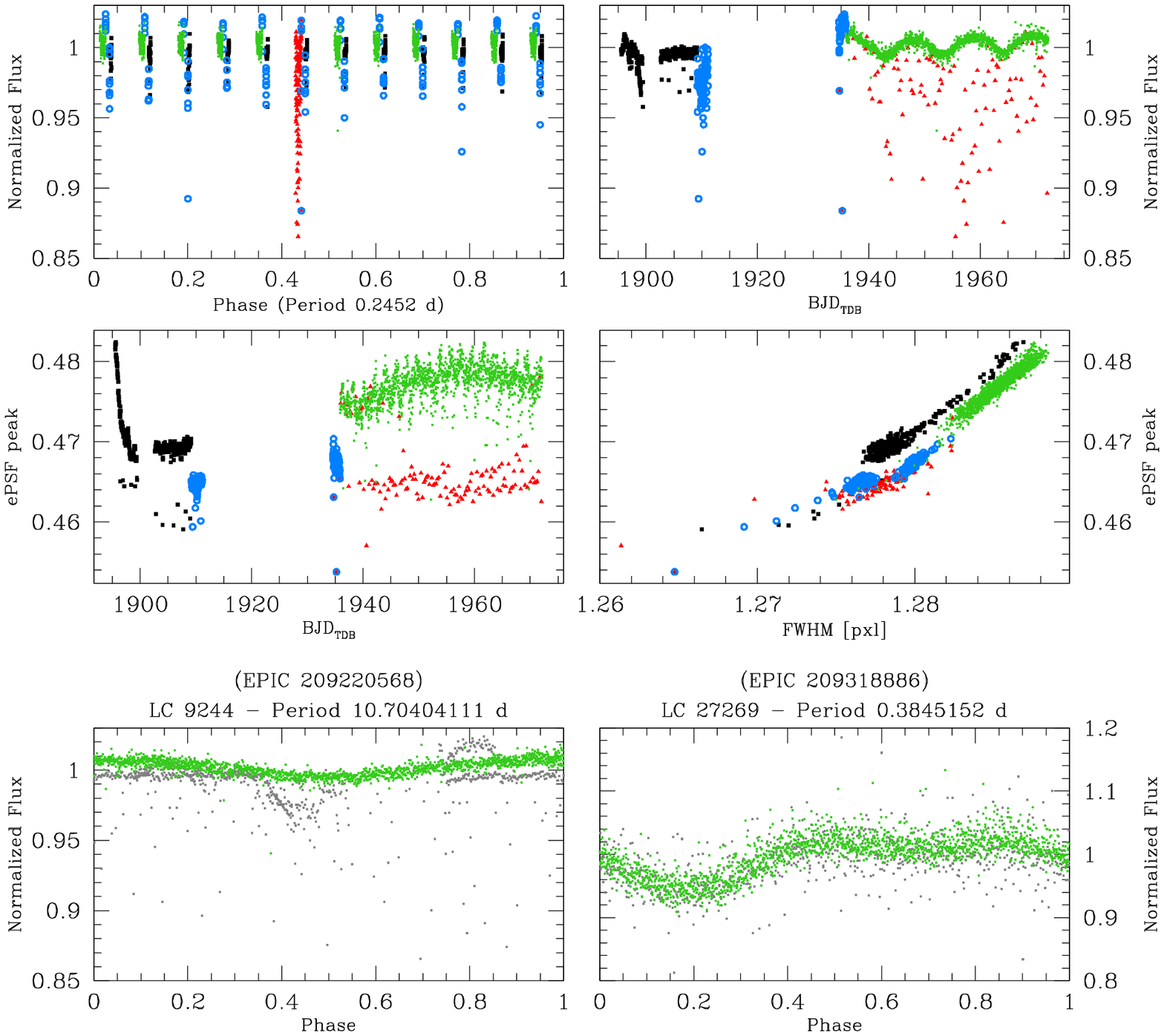}
  \caption{Clean LC definition. In the \textit{Top-left} panel, we
    show the phased LC with a period of 0.2452 days for the object \#
    9244 in the AIC. We mark with black filled squares the points
    obtained before the second safe mode during the C0 observations
    and with azure, open circles the points corresponding to the
    ``coarse pointing'' flag in the original TPFs. The red triangles
    highlight the exposures that are associated to a thruster-jet
    event. Finally, with green dots we show the best 1629 out of 2422
    points (exposures) for the LC analysis. In the \textit{Top-right}
    we show the normalized flux as a function of the time for the same
    LC, with the points colour-coded as before. In the
    \textit{Middle-left} panel we plot the peak value of the perturbed
    ePSF of each exposure as a function of time. In the ePSF peak-FWHM
    plane (\textit{Middle-right} panel) it is also clear that there is
    a correlation between every effect that could harm the
    observations and the ePSF. This is an hint of the usefulness of
    the ePSF parameters as a diagnostic tool. Finally, in the
    \textit{Bottom} panels we show the LC for the bright object
    illustrated above (\textit{Left}) and for a fainter star
    (\textit{Right}, AIC \# 27269) in which we plot with green and
    gray dots the good and the bad points, respectively.}
  \label{fig11}
\end{figure*}

In Fig.~\ref{fig11}, we show an outline of our best-exposure selection
for a variable-candidate LC (AIC star \# 9244). It is well known that
\textit{K2}/C0 data contain a periodic, systematic effect every
$\sim$6 hours, related to the thurster-jet firings used to keep the
roll-angle in position. Indeed high SNR stars show a well-defined
periodic effect every $\sim$0.2452 days (and at its harmonics). In the
top-left panel of Fig.~\ref{fig11} we show the hand-selected points in
the phased LC with a period of $\sim$0.2452 days, while in the
top-right panel we show the corresponding location of such points
during C0. By phasing different LCs with this period, we noticed that
there is one group of points (red, solid triangles) that is more
scattered than the others. These points are associated with the
thruster-jet events. The remaining points, marked with different
colours and shapes, represent different portions of the LC. The less
populated groups (black, solid squares) were taken during the first
part of the campaign; while the more populated clumps (green dots)
were taken during the second part of the C0. The remaining outliers
(azure, open circles) are those points obtained during the coarse
pointing of the spacecraft, flagged in the TPFs accordingly. In
practice, we discarded all points corresponding to the first part of
C0, and the points collected during the thruster-jet events and coarse
pointing. \\

We found very good agreement between such point selection and the ePSF
(middle-left panel) peak and FWHM (middle-right panel). For example,
the exposures taken during a thruster-jet event have a less-peaked
ePSF because the exposure is blurred during the long-cadence
integration time. Hereafter, we use the adjective ``\textit{clean}''
to define a LC based on the stable part on the second half of the
Campaign (green points in Fig.~\ref{fig11}). We run our variable star
finding algorithms only on the clean LCs.

\section{Photometric precision} 
\label{photprec}

We used three different parameters to help us to estimate the
photometric precision:

\begin{itemize}

\item \textit{rms} (defined as in Sect~\ref{detrend});

\item \textit{p2p rms}. The point-to-point rms (p2p rms) is defined as
  the 3.5$\sigma$-clipped 68.27$^{\rm th}$-percentile of the
  distribution around the median value of the scatter (the difference
  between two consecutive points);

\item \textit{6.5-hour rms}. The \textit{6.5-hour rms} is defined as
  follows. We processed each available LC with a 6.5-hour
  running-average filter, and then divided it in 13-point bins. For
  each bin, we computed the 3.5$\sigma$-clipped rms and divided it by
  $\sqrt{12}$ ($\sqrt{N-1}$, with $N$ the number of points in each
  bin). The 6.5-hour rms is the median value of these rms
  measurements.

\end{itemize} 
All three parameters have been calculated on the normalized LC.

Our PSF-based, neighbour-subtracted technique has been specifically
developed to deal with crowded regions and faint stars. Therefore, we
expect substantial improvements with respect to what is in the
literature in these two specific regimes. In the following, we will
demonstrate the effectiveness of our new approach.

\begin{figure*}
  \centering
  \includegraphics[width=\textwidth]{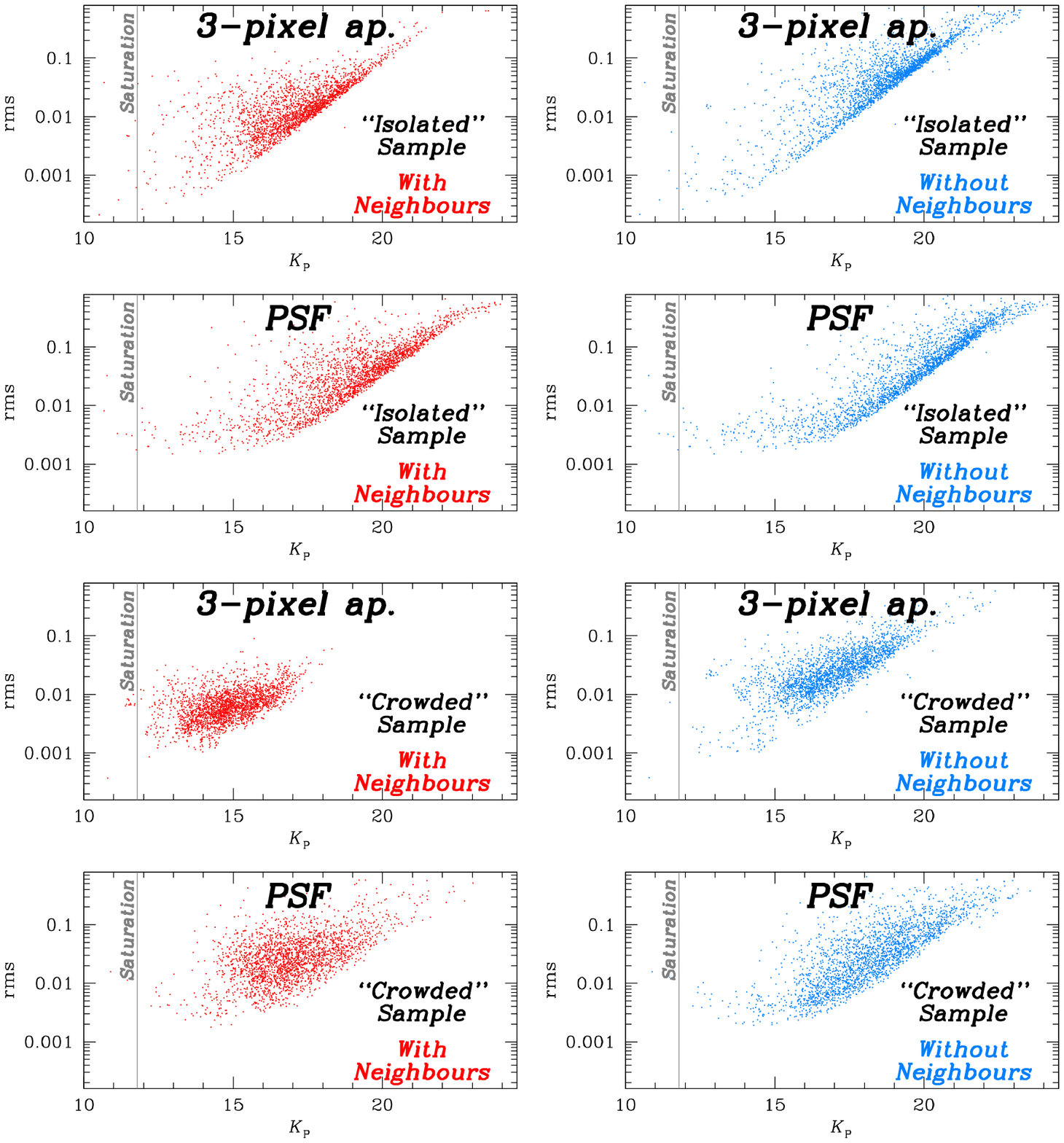}
  \caption{Photometric rms of LCs extracted from the original (red
    points in \textit{Left} panels), and the neighbour-subtracted
    (azure points in \textit{Right} panels) images as a function of
    the $K_{\rm P}$ magnitude. We show the rms trend for the
    ``isolated'' (the first and second rows from \textit{Top}) and the
    ``crowded'' (the third and fourth rows) samples, defined as
    described in the text, for LCs obtained using a 3-pixel aperture
    and the PSF. The saturation threshold is set at $K_{\rm
      P}$$\sim$11.8 (vertical, gray solid line).}
  \label{fig12}
\end{figure*}

\subsection{Photometry on images with and without neighbours}
\label{photprecwwo}

By subtracting stellar neighbours before measuring the flux of a given
star, we can obtain performance comparable with that achieved by
others in the literature, but in more crowded regions. In general,
within a single 4$\times$4-arcsec$^2$ pixel, there can be more than
one source that contribute to the total flux. Therefore, we expect the
neighbour-subtraction method to be useful not only in stellar clusters
or fields in the Galactic bulge, but also in relatively-lower-density
regions.

To demonstrate this assertion, we selected two samples of stars: a
``crowded'', sample centered on NGC~2158, and an ``isolated'' sample,
that comes from five different regions where the stellar density is
lower. The two samples contain the same number of sources. We computed
the LC rms for a 3-pixel aperture and for PSF photometry with and
without subtracting stellar neighbours (Fig.~\ref{fig12}).

Without removing stellar neighbours, the light of close-by stars that
falls within the aperture increases the target flux, alters the
faint-end tail of the rms trend (naturally, the problems becomes
larger for larger apertures). The effect is more evident in crowded
regions, in particular when using aperture photometry. For the 3-pixel
aperture photometry, in the crowded region the limiting magnitude on
the original images is $K_{\rm P}\sim$18, to be compared with the
$K_{\rm P}\sim$22 limiting magnitude in neighbour-subtracted images of
the same field (second-to-the-last panels in Fig.~\ref{fig12}). In
summary, in crowded regions, by using the neighbour-subtracted images,
we obtained a more reliable stellar flux and gained about 4 $K_{\rm
  P}$ magnitudes for the 3-pixel-aperture photometry. Furthermore, for
both aperture and PSF photometry, we have a lower rms for the LCs, and
the bulk of the LC-rms distribution as a function of magnitude looks
sharper. In conclusion, hereafter we will consider only
neighbour-subtracted LCs.

\subsection{Photometry on bright and faint stars}
\label{photprec3psf}

As expected, and as shown by \cite{Nar15} for ground-based data,
aperture photometry performs, on average, better on isolated, bright
stars, while the PSF photometry gives better results on faint
stars. In Table~\ref{tab:rms1} and \ref{tab:rms2} we list the rms
values in part-per-million (ppm) for each of the five photometric
methods we adopted.

As can be inferred from these Tables, for bright stars,
3-pixel-aperture photometry shows, on average, lower rms than 5- and
10-pixel aperture photometry. The 1-pixel-aperture- and the PSF-based
photometry have almost the same trend in the faint star regime
($K_p\ge 15.5$), with the 1-pixel aperture showing a slightly smaller
rms for $K_{\rm P}$$>$19, while the PSF photometry performs 2-3 times
better on brighter stars. In the following discussion, for faint
stars, we prefer to use PSF rather than 1-pixel-aperture
neighbour-subtracted photometry, although, on average, equivalent in
terms of photometric scatter.

\begin{figure*}
  \centering
  \includegraphics[width=\textwidth]{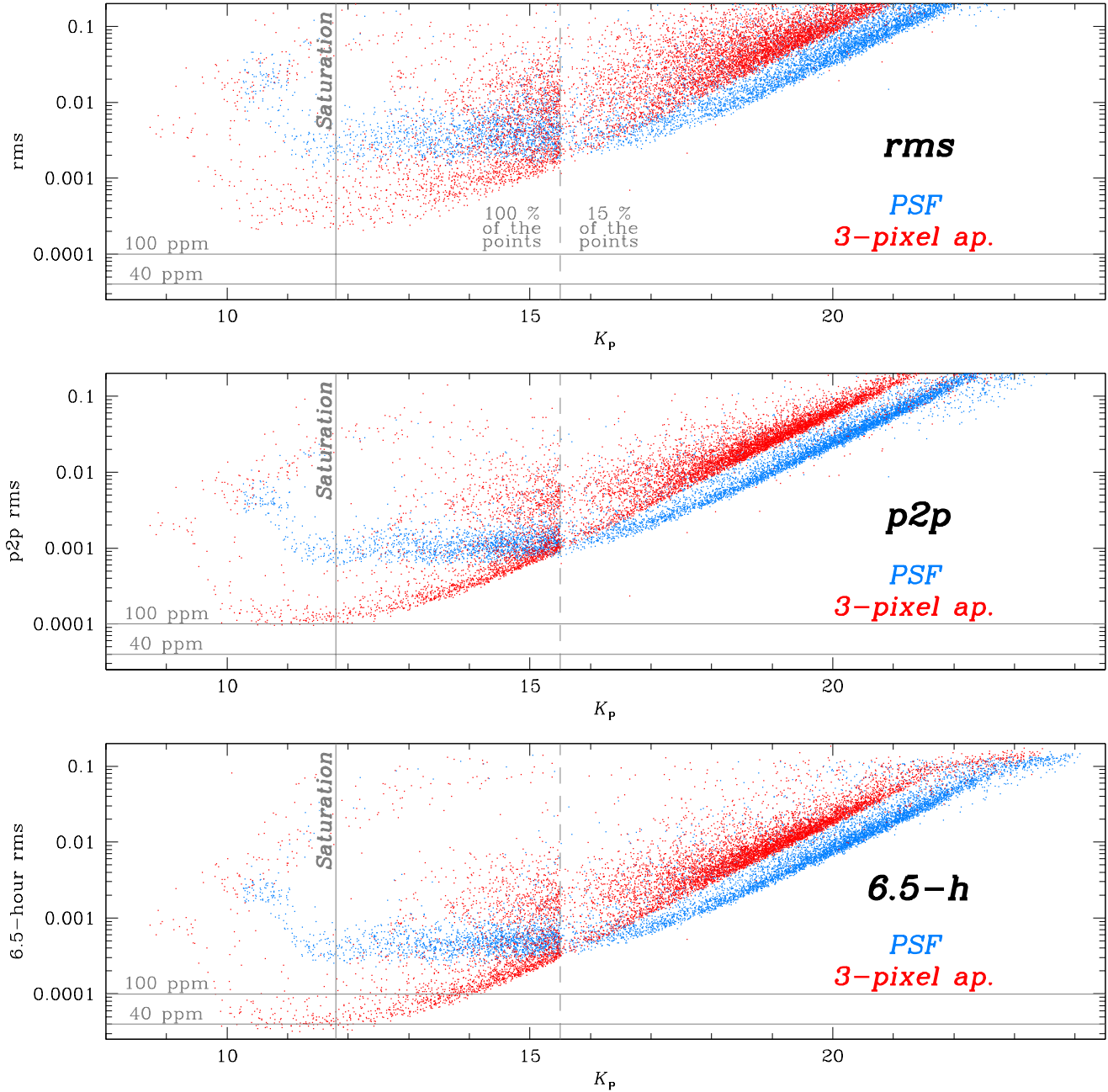}
  \caption{Photometric rms (\textit{Top}), p2p rms (\textit{Middle})
    and 6.5-hour rms (\textit{Bottom}) as a function of the $K_{\rm
      P}$ magnitude derived from the 3-pixel-aperture- (red points)
    and PSF-based (azure points) neighbour-subtracted LCs. The
    vertical, gray dashed line is set at $K_{\rm P}$$=$15.5. For stars
    fainter than $K_{\rm P}$$\sim$15.5 the PSF photometry has a lower
    rms. The vertical, gray solid line is set at the saturation
    threshold ($K_{\rm P}$$\sim$11.8). As a reference, we plot two
    horizontal, gray solid lines at 100 and 40 parts-per-million
    (ppm), respectively. As in Fig.~\ref{fig11}, for $K_{\rm
      P}$$>$15.5 we show only the 15\% of the points, for clarity.}
  \label{fig13}
\end{figure*}

In Fig.~\ref{fig13} we show a comparison between the PSF and the
3-pixel aperture photometric methods. It is clear that for $K_{\rm
  P}$$>$15.5 the PSF photometry performs better than the 3-pixel
aperture.

For bright stars, the 3-pixel-aperture 6.5-hour rms is below 100 ppm
between 10$<$$K_{\rm P}$$<$14, with a best value of about 30 ppm for
stars with 11$<$$K_{\rm P}$$<$12.5. For stars brighter than $K_{\rm
  P}$$\sim$10, the rms increases, mainly because we are working on
heavily-saturated stars. In the faint-end regime, using PSF photometry
we obtained 6.5-hour rms 2-3 times better than using a
3-pixel-aperture photometry. At 18$<$$K_{\rm P}$$<$19 the 6.5-hour rms
is about 2600 ppm. Such precision allows us to detect a flux drop of a
few hundredths of magnitude, i.e., the LC dimming due to exoplanet
candidate TR1 discovered by \cite{Moc06}. We will further discuss this
object in Sect.~\ref{TR1sect}.

In conclusion, we find that even at the bright end our
neighbour-subtracted technique allows us to obtain performance that is
comparable to those in the literature. This is the case even though we
are dealing with a more crowded region than most previous studies.

\subsection{Comparison with existing works on \textit{K2} data}
\label{rmslit}

At present, there are a number of studies in literature that are
focused on the Campaign 0 data of \textit{K2}. Here, we provide a
comparison, both using the LC rms and through a visual inspection of
the LCs, for the objects in common with previous published studies.

Before making such a comparison, we warn the reader about some aspects
that should be taken into account in the following. A fair comparison
should be made by comparing the single light curves point by
point. Unfortunately, thus far, no one has ever attempted to measure
all stars in the entire super-stamps. Rather it is more common
practice to restrict the analysis to the outer parts of dense stamps,
where the crowding is less severe.

That said, we do have several LCs from stars in common with various
studies; in the best case \citep{V&J14}, we have 40
stars. Furthermore, by simply comparing the rms given in the different
papers, we could introduce some biases. For example, the rms in a
given magnitude bin can be overestimated by the number of variable
stars within that bin; the methods adopted to compute the rms can be
different; there might be different calibration methods to transform
``raw'' magnitudes into the \textit{Kepler} photometric system. In the
latter case, stars can fall in different magnitude bins in different
papers. Finally, our neighbour-subtracted LCs are less affected by
neighbour-light contamination. As noticed above, the light pollution
would result in a brighter LC, which would move the star into a
brighter magnitude bin, and decrease its rms (because of the higher
number of counted photons).

\begin{table*}
  \caption{Photometric precision of the 3-pixel-aperture- and
    PSF-based photometry evaluated as described in
    Sect.~\ref{photprec}. The values are given in part-per-million. We
    used the clean LCs to compute these quantities. When no stars were
    found in a given magnitude interval, we inserted a ``/'' in the
    corresponding cell.}  \centering
  \label{tab:rms1}
  \begin{tabular}{ccccccc}
    \hline
    \hline
    $K_{\rm P}$ Magnitude & \multicolumn{3}{c}{3-pixel aperture} & \multicolumn{3}{c}{PSF} \\ 
    interval  & rms & p2p & 6.5-h rms &  rms & p2p & 6.5-h rms \\
    \hline 
     8 -     9 &       4365 &       1994 &        840 &          / &          / &          / \\ 
     9 -    10 &       4550 &       1915 &        771 &          / &          / &          / \\ 
    10 -    11 &       3083 &        433 &        101 &      18231 &       3918 &       2054 \\ 
    11 -    12 &        446 &        124 &         44 &       2636 &        949 &        424 \\ 
    12 -    13 &        744 &        168 &         62 &       3300 &       1068 &        459 \\ 
    13 -    14 &        926 &        276 &         97 &       3200 &        983 &        439 \\ 
    14 -    15 &       1977 &        585 &        196 &       3871 &       1085 &        489 \\ 
    15 -    16 &       5720 &       1882 &        633 &       3988 &       1274 &        528 \\ 
    16 -    17 &      10631 &       4731 &       1606 &       4822 &       1875 &        693 \\ 
    17 -    18 &      21219 &       9533 &       3176 &       7214 &       3518 &       1189 \\ 
    18 -    19 &      40885 &      19749 &       6564 &      14070 &       7936 &       2601 \\ 
    19 -    20 &      76625 &      41163 &      13534 &      32813 &      18376 &       6150 \\ 
    20 -    21 &     159131 &      89546 &      29876 &      71199 &      40570 &      13456 \\ 
    21 -    22 &     327174 &     179254 &      71173 &     153875 &      89963 &      30335 \\ 
    22 -    23 &     487721 &     282520 &     109300 &     309403 &     179357 &      73297 \\ 
    23 -    24 &     606784 &     412491 &     140349 &     468002 &     292661 &     114074 \\ 
    \hline
  \end{tabular}
\end{table*}

\begin{table*}
  \caption{As in Table~\ref{tab:rms1}, but for 1-, 5- and
    10-pixel-aperture-based photometry.}
  \centering
  \label{tab:rms2}
  \begin{tabular}{c|ccc|ccc|ccc}
    \hline
    \hline
    $K_{\rm P}$ Magnitude & \multicolumn{3}{c}{1-pixel aperture} & \multicolumn{3}{c}{5-pixel aperture} & \multicolumn{3}{c}{10-pixel aperture} \\ 
    interval  & rms & p2p & 6.5-h rms &  rms & p2p & 6.5-h rms & rms & p2p & 6.5-h rms \\
    \hline
     7 -     8 &          / &          / &          / &          / &          / &          / &       5222 &       2032 &        917 \\ 
     8 -     9 &          / &          / &          / &       3983 &       1758 &        802 &       3640 &       1230 &        582 \\ 
     9 -    10 &          / &          / &          / &       8641 &       2982 &       1558 &       6083 &       2311 &       1152 \\ 
    10 -    11 &       9126 &       5411 &       1755 &      10449 &       4554 &       1745 &      18628 &       7103 &       3199 \\ 
    11 -    12 &       5625 &       3552 &       1122 &        451 &        116 &         43 &       1397 &        261 &         87 \\ 
    12 -    13 &       5594 &       3461 &       1098 &       1023 &        246 &         87 &       2972 &       1081 &        437 \\ 
    13 -    14 &       5589 &       3482 &       1113 &       3959 &        989 &        387 &      10915 &       3876 &       1350 \\ 
    14 -    15 &       5918 &       3484 &       1115 &       7777 &       2897 &       1043 &      15341 &       6028 &       2083 \\ 
    15 -    16 &       6108 &       3566 &       1141 &      13369 &       5306 &       1866 &      24145 &      10945 &       3738 \\ 
    16 -    17 &       6473 &       3743 &       1201 &      21162 &       8917 &       3056 &      44430 &      21238 &       7173 \\ 
    17 -    18 &       8003 &       4477 &       1439 &      39276 &      17885 &       6022 &      84555 &      43223 &      14664 \\ 
    18 -    19 &      13766 &       7833 &       2521 &      69978 &      36431 &      12105 &     170644 &      91746 &      34174 \\ 
    19 -    20 &      30515 &      17248 &       5557 &     141595 &      77484 &      26331 &     285471 &     134808 &      69279 \\ 
    20 -    21 &      66356 &      37814 &      12241 &     299378 &     158724 &      63791 &     404477 &     185638 &      87072 \\ 
    21 -    22 &     141837 &      83477 &      27306 &     432749 &     225126 &      97786 &     528062 &     289481 &     116628 \\ 
    22 -    23 &     301746 &     173745 &      66720 &     591510 &     359285 &     128364 &     587527 &     373223 &     138088 \\ 
    23 -    24 &     447170 &     287843 &     106035 &     687907 &     511448 &     147173 &          / &          / &          / \\
    \hline
  \end{tabular} 
\end{table*}

\begin{figure*}
  \centering
  \includegraphics[width=\columnwidth]{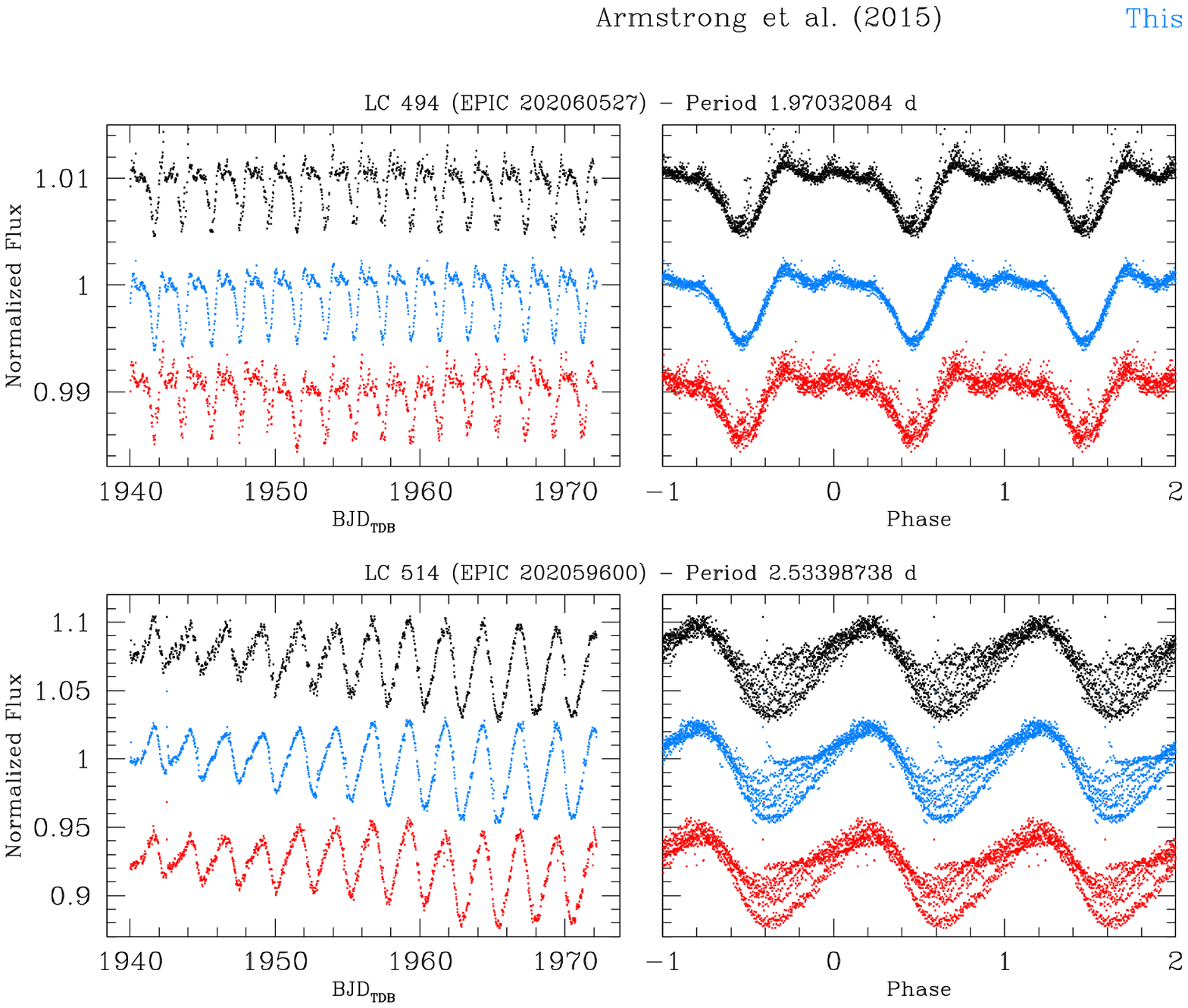}
  \includegraphics[width=\columnwidth]{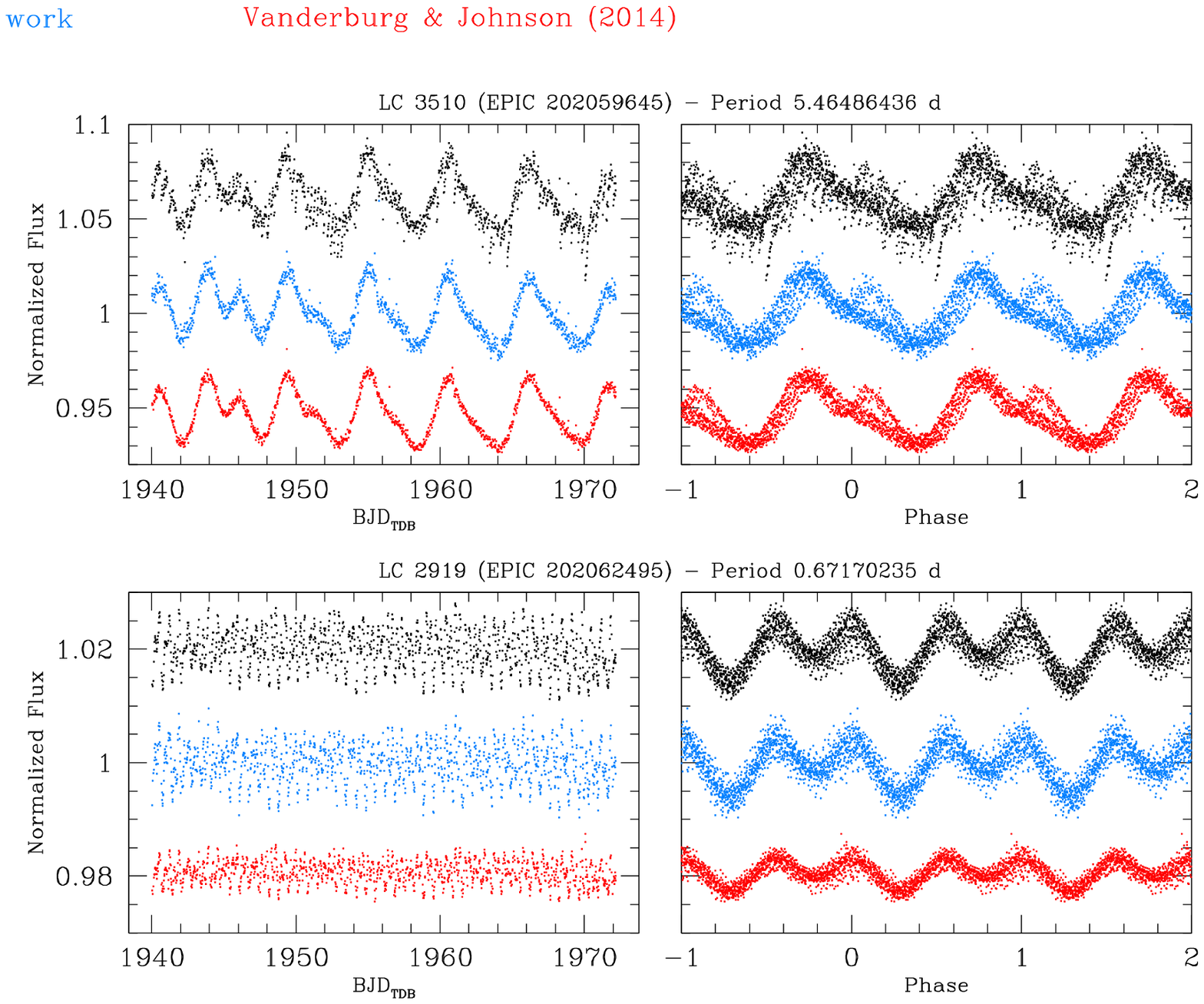}
  \vskip 5 pt
  \includegraphics[width=\columnwidth]{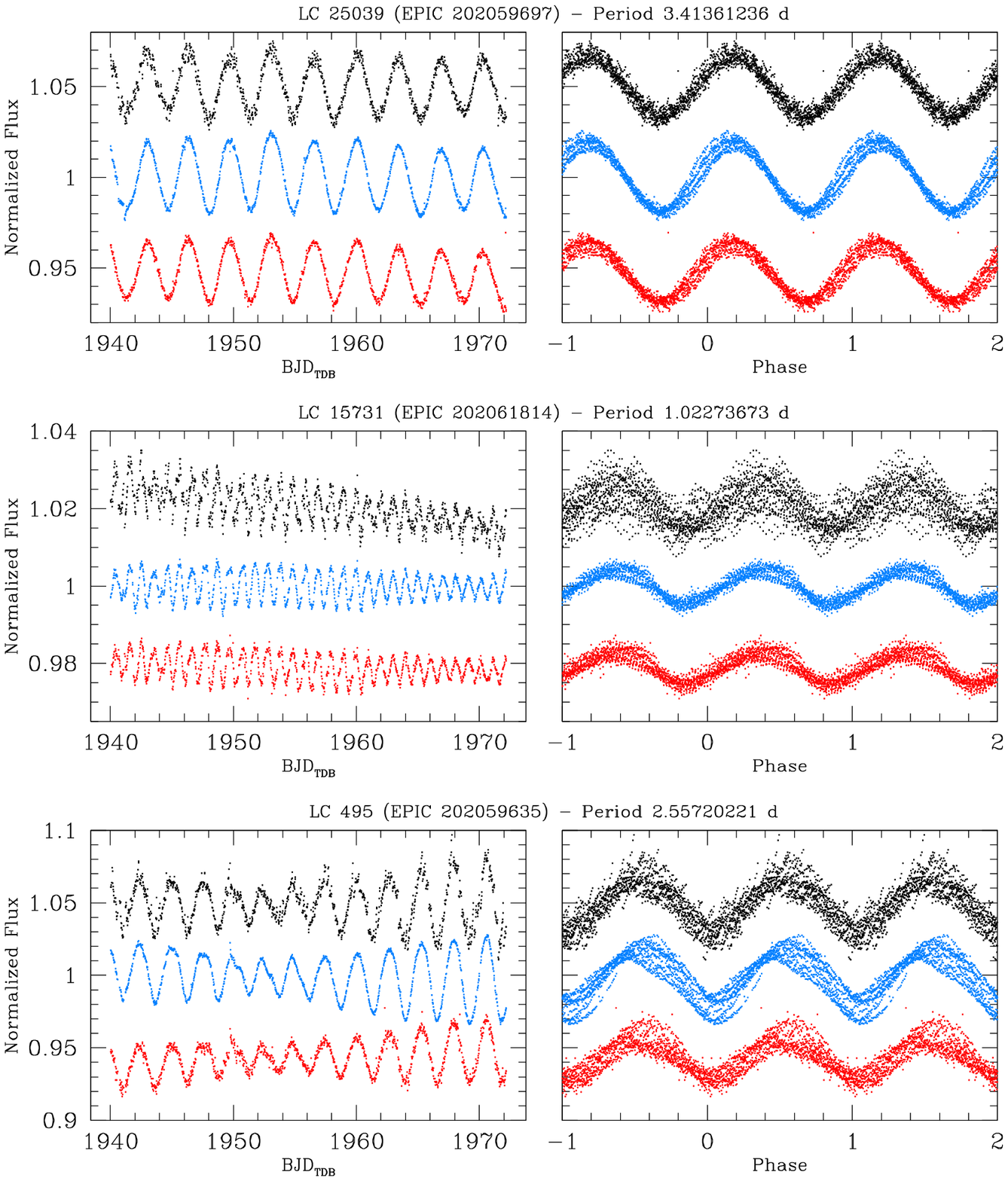}
  \includegraphics[width=\columnwidth]{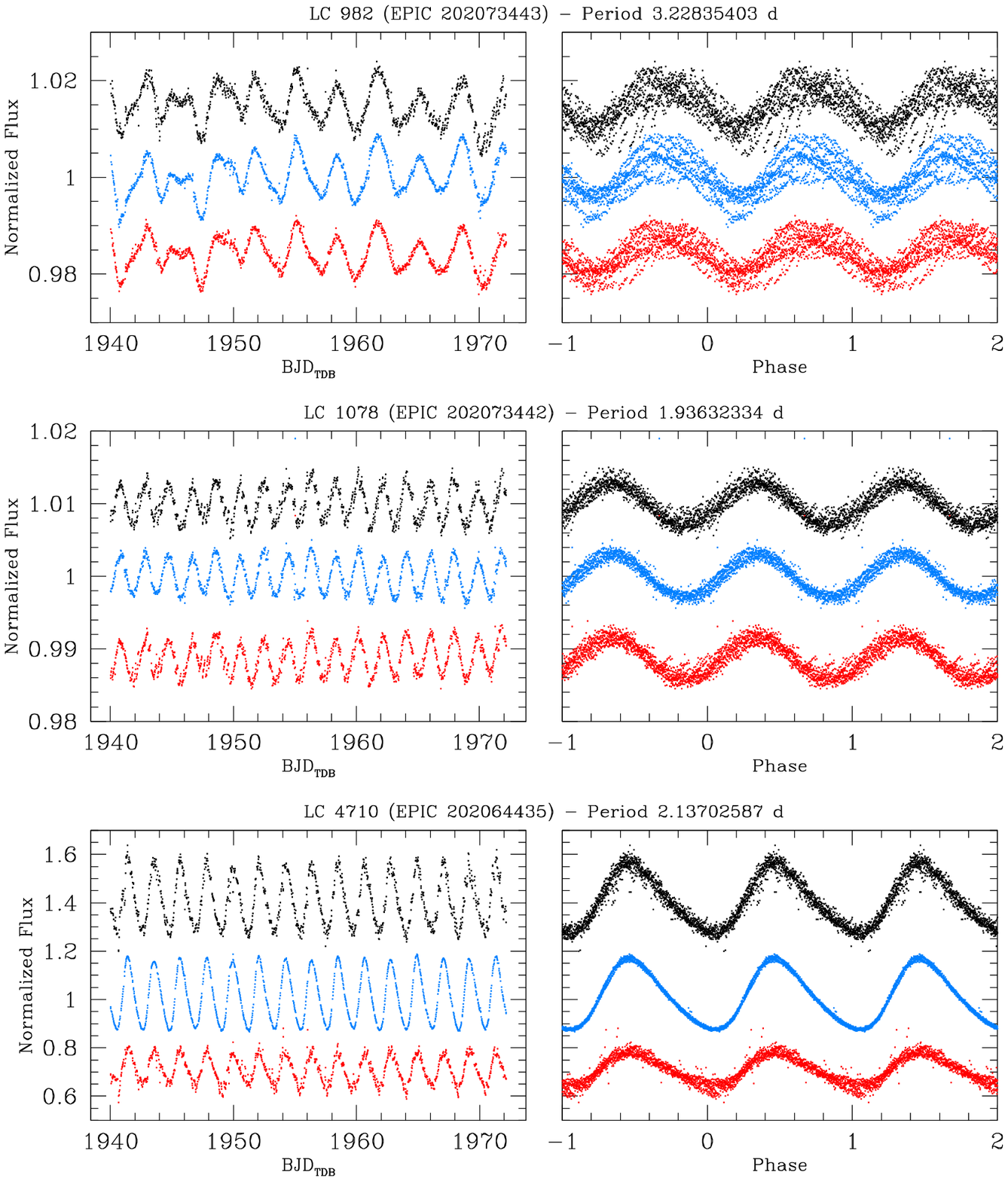}
  \caption{Point-by-point comparison between our light curves (azure
    points) and those from \citet{V&J14} (red) and \citet{Arm15}
    (black). We show only the points imaged at the same BJD$_{\rm
      TDB}$ in common among the three works.}
  \label{fig14}
\end{figure*}

\subsubsection{Comparison with \citet{V&J14}}

The first published study on \textit{K2} photometry was that of
\cite{V&J14}, and was based on the Engineering data set. The Campaign
0 data set was subsequently reanalyzed by \cite{Van14}, who worked to
improve the photometric performance for fainter stars.

We downloaded from the MAST archive all his LCs for the targets within
channel 81, and found 40 objects in common with our catalogue.  We
computed the three rms as described above. The LCs from \cite{Van14}
show, on average, a larger p2p and 6.5-hour rms than our LCs for the
variable objects, but their p2p and 6.5-hour rms for non-variable
stars was lower. While the former behavior was predictable, (as
reported by \citealt{V&J14}, their self-calibrated-flat-field approach
works best on dwarfs rather than on highly-variable stars) the latter
trend was unexpected. Therefore, we also visually inspected the
location of all stars in common on the images and the regions covered
by the adopted aperture
masks\footnote{\href{https://www.cfa.harvard.edu/~avanderb/k2.html}{https://www.cfa.harvard.edu/$\sim$avanderb/k2.html}}.

For bright stars, \citet{Van14} used a circular aperture that included
several neighbour sources. For faint stars, the ad-hoc aperture that
was designed to avoid light contamination works only partially, since
the flux on the wings of the PSF can still fall inside
it. Furthermore, for blended stars the adopted aperture mask included
all of them and, finally, the EPIC catalog seems not complete
enough. In either cases, the total flux of the source is increased
because of the contribution of the neighbours, resulting in an higher
SNR for the source (with a consequent better Poisson rms), but this
does not correspond to the true SNR of the individual target
source. Of course, we cannot exclude that \cite{Van14} detrending
algorithm works better, further improving the final rms. A larger
sample of stars is required for a more conclusive comparison.

We also found that at least two variable objects identified by
\cite{Van14} were mismatched/blended, difficult for him to identify
due to the limitations introduced by the EPIC (bright, isolated
objects) and to the low resolution of the \textit{K2} images (see next
Section).

It is worth to mention that \cite{Van14} releases for each light curve
a flag that marks the cadence number associated to a thruster-jet
event. We compared their flag with our flag described in
Sect.~\ref{PSFrole}. Among the cadences in common between the two
works, we found that we flagged (and do not used) a lower number of
images. This is probably due to our perturbed-PSF approach that is
able to fit reasonably well objects imaged while the thruster-jet
effect was not severe. However, we found a good agreement between the
two thruster-jet-identification methods.

\subsubsection{Comparison with \citet{Arm15}}

\cite{Arm15} released C0 LCs obtained with a similar method and with
comparable performance to those of \cite{V&J14}. We downloaded the LCs
from their
archive\footnote{\href{http://deneb.astro.warwick.ac.uk/phrlbj/k2varcat/}{http://deneb.astro.warwick.ac.uk/phrlbj/k2varcat/}}
and computed the rms for the 12 objects we have in common. Again, we
found that our photometry provides LCs with a lower rms for almost all
these variable stars.

As the rms cannot give a direct measurement of the goodness of the
photometry, we made a visual comparison of the 12 LCs in common
between our data set and \cite{Arm15} one.  In Fig.~\ref{fig14} we
show the LC comparison of 10 out of 12 objects in common. We also
included the LCs of \cite{V&J14} since these objects are present in
their sample too. We plotted the best photometry for all the LCs. On
average, our LCs looks sharper (e.g., LC 4710).

\begin{figure}
  \centering
  \includegraphics[width=\columnwidth]{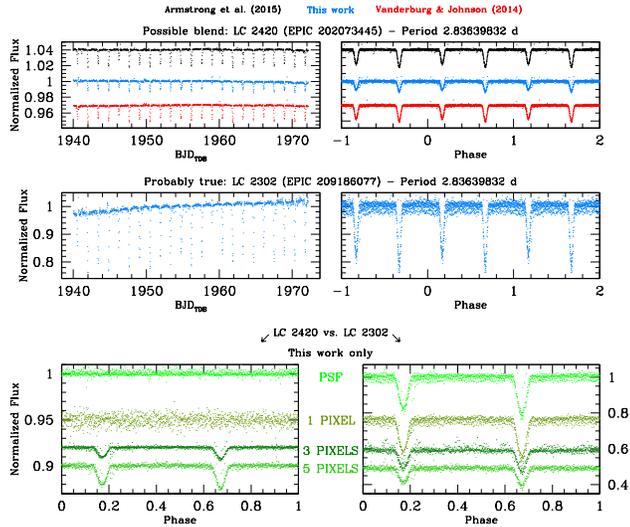}
  \caption{LC comparison between EPIC~202073445 (LC \# 2420 in AIC)
    and EPIC~209186077 (LC \# 2302). On the \textit{Top} panels we
    compared our LC of the probable-blend object (EPIC~202073445) with
    those found in the literature as in Fig.~\ref{fig14}. In the
    \textit{Middle} panels we show the LC of the object that we found
    to be the true variable (EPIC~209186077). We show only our LC
    since these objects are neither in \citet{V&J14} nor in
    \citet{Arm15} data set. In the \textit{Bottom} panels we finally
    show our LCs colour-coded with different shades of green according
    to the photometric method (PSF, 1-, 3- and 5-pixel aperture).}
  \label{fig15}
\end{figure}

Fig.~\ref{fig15} shows the first of the remaining two objects we have
in common. EPIC~202073445 is an eclipsing binary. The depth of our LC
is smaller than in the other two papers. However, \cite{Nar15} found
that the real eclipsing binary is another star (namely,
EPIC~209186077), very close to EPIC~202073445. To further shed light
on this ambiguity, we compared the LCs of both stars using different
kinds of photometry. The mismatched eclipsing binary does not show any
flux variation with the PSF and 1-pixel-aperture photometry. On the
other hand, the larger the aperture, the deeper the eclipse. Instead,
the ``real'' eclipsing binary shows that the eclipses become dimmer
and dimmer with increasing aperture radius, as expected since its flux
is diluted by the remaining flux of the (un-)subtracted neighbours. We
confirm that the true eclipsing binary is the one identified by
\cite{Nar15} (EPIC~209186077). This mis-identification is also present
in the eclipsing-binary catalogue of \cite{LaC15}.

\begin{figure}
  \centering
  \includegraphics[width=\columnwidth]{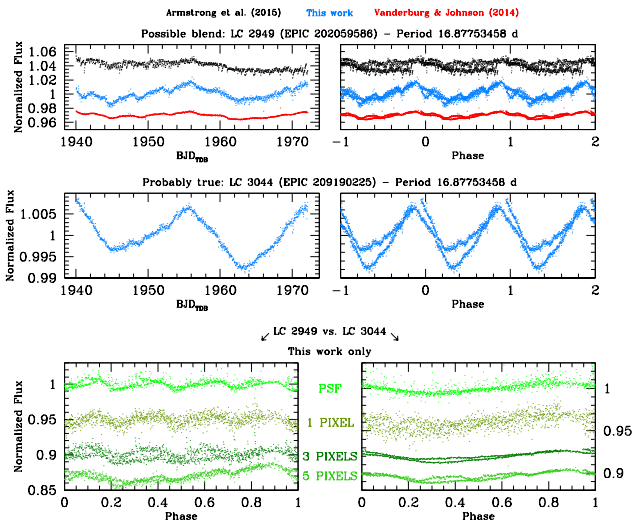}
  \caption{As in Fig.~\ref{fig15}, but for the case of
    EPIC~202059586 (LC \# 2949) and EPIC~209190225 (LC \# 3044).}
  \label{fig16}
\end{figure}

There is a similar ambiguity between EPIC~202059586 and EPIC~209190225
(Fig.~\ref{fig16}). Again, comparing the PSF- and aperture-based LCs,
we found that the first object is rather an aperiodic star that shows
a flux modulation due to the latter object, while the second one is
the real variable. \\

\subsubsection{Comparison with \citet{Aig15}}

\cite{Aig15} developed a different method, which is described in
detail in their paper. Their approach has some similarities with the
effort here described, e.g., the image reconstruction and the adoption
of an input list. Unfortunately, a proper comparison is not possible
since they analyzed only the Engineering data. Campaign 0 showed
different problems and lasted longer. Anyway, for the sake of
completeness, we compared their results with our LCs.

Due to its less-ambiguous definition, we compared the p2p rms. In the
brightest magnitude intervals (9$<$$K_{\rm P}$$<$11), \cite{Aig15}
shows a better p2p rms than we derived here. In the interval
11$<$$K_{\rm P}$$<$15, our 3-pixel-aperture p2p rms ranges from 124 to
585 ppm; while their p2p rms varies between 238 and 867 ppm. For
fainter magnitudes up to $K_{\rm P}$$\sim$19 our p2p rms is slightly
better. On the other hand, our PSF-based photometry performs much
better in the magnitude interval 15$<$$K_{\rm P}$$<$19, with a minimum
p2p-rms value of 1274 ppm and a maximum of 7936 ppm, to be compared
with that of \cite{Aig15} that increases from 1841 to 23673 ppm in the
same magnitude range. It is worth to mention that we could measure
objects up to 5 magnitudes fainter than those measured by
\cite{Aig15}.

\begin{figure}
  \centering
  \includegraphics[width=\columnwidth]{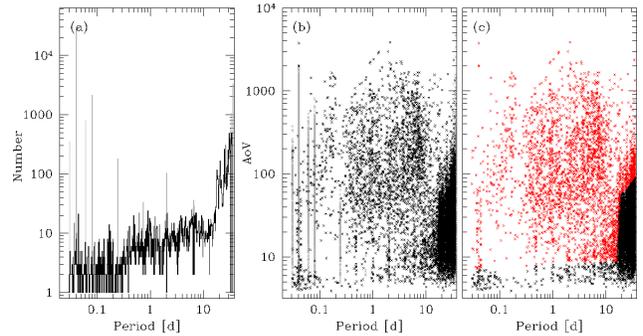}
  \caption{Example of candidate-variable selection using the AoV
    algorithm.  Panel (a): the distribution of the periods; in black
    we show the distribution after removing the spikes, in grey the
    removed spikes.  Panel (b): LC periods as a function of the AoV
    metric $\Theta$ for all the stars; the stars that have passed the
    first selection are plotted in black, the stars having periods
    corresponding to the removed spikes in gray.  Panel (c): periods
    of light curves as a function of $\Theta$ after spikes
    suppression. Red dots represent variable candidates.}
  \label{fig17}
\end{figure}

\section{Variable Candidates}

\subsection{Search for Variables} 
\label{search}

In order to detect candidate variable stars, we followed the procedure
adopted by \cite{Nar15}; we describe the basics of the procedure
below. We obtained the periodograms of all clean LCs using three
different tools included in the \texttt{\small VARTOOLS} v1.32
package, written by \cite{Hart08} and public
available\footnote{\href{http://www.astro.princeton.edu/$\sim$jhartman/vartools.html}{http://www.astro.princeton.edu/$\sim$jhartman/vartools.html}}:

\begin{itemize}

\item The first tool is the Generalized Lomb-Scargle (GLS) periodogram
  (\citealt{Press92}; \citealt{Zech09}), useful in detecting
  sinusoidal periodic signals. It provides the formal false alarm
  probability (FAP) that we used to select variable-star candidates.

\item The second tool is the Analysis of Variance (AoV) periodogram
  (\citealt{SC89}), suitable for all kind of variables. We used the
  associated AoV FAP metric ($\Theta$) that is a good diagnostic
  useful to select stars that have a high probability to be variables.

\item The third tool is the Box-fitting Least-Squares (BLS)
  periodogram (\citealt{KZM02}), particularly effective when searching
  for box-like dips in an otherwise-flat or nearly-flat light curve,
  such as those typical of detached eclipsing binaries and planetary
  transits.  We used the diagnostic ``signal-to-pink noise''
  (\citealt{Pont06}), as defined by \cite{Hart09}, to select
  eclipsing-binary and planetary-transit candidates.

\end{itemize}

Figure~\ref{fig17} illustrates the procedure used to identify
candidate variables resulting by the application of the AoV finding
algorithm.  The same procedure has been adopted for the GLS and BLS
finding algorithms. First, we built the histogram of the periods ($P$)
of all clean LCs, as shown in panel (a) of Fig.~\ref{fig17}. For each
period $P_0$ we computed the median of the histogram values in the
bins within an interval centered at $P_0$ and extended by $50 \times
\delta P$, where $\delta P$ is the bin width chosen to build the
histogram.  We flagged as ``spike'' the $P_0$ corresponding to a
histogram value $5\sigma$ above that median, where $\sigma$ is the
68.27$^{\rm th}$ percentile of the sorted residuals from the median
value itself.  These spikes are associated with spurious signals due
to systematic effects such as, e.g., the jet-firing every $\sim$5.88
hours, the long cadence at $\sim$29 minutes ($\sim$0.02 d) and the
periodicity of $\sim$2.04 days (described in Sect.~\ref{204d}) and
their harmonics. Finally, we removed from the catalogue the stars
having periods inside $P_0 \pm \delta P/2 $, keeping only those with
high $\Theta$. We performed our searching to periods between 0.025
(slightly higher than the long-cadence sampling) and 36.5 d (clean-LC
total time interval). GLS, AOV and BLS input parameters were chosen to
optimize the variable finding and are not perfect. Indeed, we selected
a sample of known variables and tuned the input parameters in order to
maximize the signal-to-noise ratio outputted by each of the three
findings. This way, some variables could have been missed because the
three tasks found the highest SNR with a wrong period, e.g., the
0.04-d alias.

In panel (b) of Fig.~\ref{fig17}, for all clean LCs, we plot the
$\Theta$ parameter as a function of the detected period, highlighting
the objects removed because their period coincides with a spike. In
panel (c) we selected by hand the stars that have high $\Theta$.

We ran the \texttt{\small VARTOOLS} algorithms GLS, AoV, and BLS on a
list of 52\,596 light curves. We initially excluded all objects that
in our input list where outside the \textit{K2} TPFs. Then, for each
star, we selected the photometry (1-, 3-, 5-, 10-pixel aperture or
PSF) that gives the best precision at a given magnitude, according to
what was described in Section~\ref{photprec}, Fig.~\ref{fig13}, and
Tables~\ref{tab:rms1} and \ref{tab:rms2}.

We combined the lists of candidate variables obtained by applying the
three variable-detection algorithms and visually inspected each of
them. We excluded all obvious blends by looking at the LC-shape and
position of each star and of its neighbours within a radius of about
11 \textit{K2} pixels ($\sim$43 arcsec).

We found a total of 2759 variables of which 1887 passed our visual
inspection as candidates and 202 were flagged as blends. The remaining
670 LCs were difficult to visually judge. We included them into our
final catalogue, but added a warning flag indicating that their
variable nature is in doubt.

\begin{figure}
  \centering
  \includegraphics[width=\columnwidth]{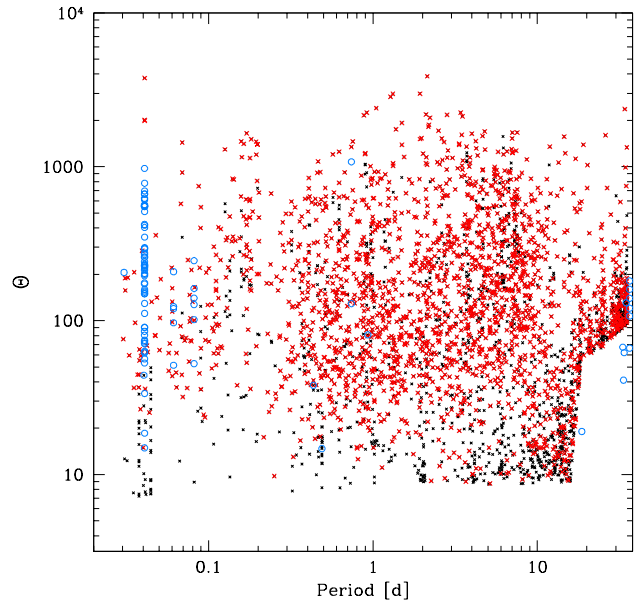}
  \caption{Periods of the light curves as a function of $\Theta$ for
    all the candidate variables. With red crosses we show all
    variables we found in our analysis of the \textit{K2} data, with
    azure, open circles the variable stars from published catalogues
    and present in our AIC we failed to identify among our \textit{K2}
    LCs. All these variables are included in our final catalogue of
    variable candidates, with different flags. Black crosses represent
    all variable sources that passes our first selection described in
    Fig.~\ref{fig17} but not our visual inspection (see text for
    detail).}
  \label{fig18}
\end{figure}

\subsection{Comparison with the literature and sample improvement}
\label{literature}

In order to evaluate the completeness of our sample, we matched the
AIC with six catalogues found in the literature: \cite{Hu05},
\cite{Jeon10}, \cite{Kim04}, \cite{Meib09}, \cite{Moc04,Moc06} and
\cite{Nar15}. These studies, which all cover the M\,35/NGC~2158
super-stamp region, analyzed different aspects of stellar variability
and made use of different observational instruments and detection
techniques (e.g., \citealt{Meib09}).

Cross-correlation of these catalogue results in 658 common entries. Of
these, 555 sources are also present in our AIC. The remaining 103
objects are missing for various reasons. A small fraction ($\sim$10\%)
of stars had a very-bright neighbour source that was badly subtracted
because the PSF is still far from perfect or were just too close to
the edge of our field of view. For the remaining missing known
variables, we found that the periods given by the three
\texttt{VARTOOLS} algorithms were close to that of a spurious signal
(spikes in the AoV vs. period plot of Fig.~\ref{fig17}) or lied below
our selection threshold (panel c in Fig.~\ref{fig17}). Therefore they
were excluded before the visual check, even if their LCs showed a
clear variable signature. As an example, we show in Fig.~\ref{fig18}
the AoV parameter $\Theta$ as a function of LC period for all the
candidate variables\footnote{Irregular and long-period objects, such
  as the cataclysmic variable V57 found by \cite{Moc04} (period
  $\sim$48 days) for which we can see only one peak in the clean LC,
  are plotted with an arbitrary period $\sim$36.5 days.} in which we
marked with a different colour the location of such missed
objects. Since there is no reason to exclude them, we added such
previously-discovered variables to our catalogue.

The variable stars found in the literature were also useful to refine
our sample and remove some blends and fake detection left after the
visual inspection of the LCs. Indeed, previous published studies have
made use of images with a higher resolution than that of \textit{K2},
and therefore we chose to rely on the former in ambiguous cases.

After this second refinement of our catalogue, we have 2133 candidate
variables, 444 sources with possible blends or dominated by systematic
effects, and 272 objects for which the LC is difficult to
interpret. In the final catalogue that we release with this paper, we
will properly flag all these different sources (see
Sect.~\ref{electronic}). In any case, we will release all LCs
extracted from the \textit{K2}/C0/channel-81 data, which will be
available to anyone for any further investigation. In
Appendix~\ref{appendixA} we show 10 LCs as example.

\subsection{Variable location on the M35 and NGC~2158 colour-magnitude diagrams}

\begin{figure*}
  \centering
  \includegraphics[width=\textwidth]{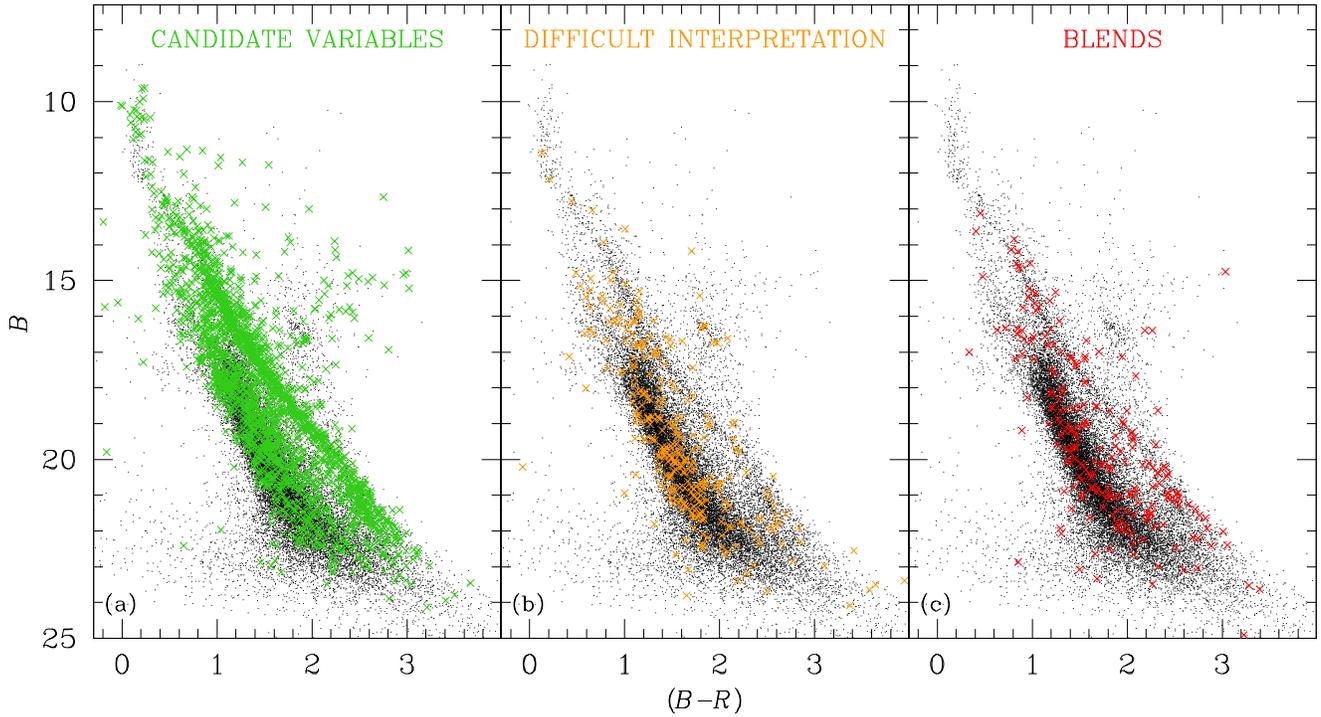}
  \caption{Distribution of our 2849 candidate variables in the $B$
    vs. ($B-R$) CMD. In the three panels we separately show the
    likely-variable stars (green crosses in panel a), the objects of
    which the LC was of difficult interpretation (orange crosses in
    panel b) and the blends (red crosses in panel c).}
  \label{fig19}
\end{figure*}

We also used the $B$$V$$R$$J_{\rm 2MASS}$$H_{\rm 2MASS}$$K_{\rm
  2MASS}$ and white-light-magnitude catalogue of \cite{Nar15} to find
the location of our variables in the colour-magnitude diagrams
(CMDs). In Fig.~\ref{fig19}, we show the $B$ vs. ($B-R$) CMDs of the
star sample used to search for variable candidates
(Sect.~\ref{search}) that have a $B$- and $R$-magnitude entry in the
catalogue. We plot candidate variables, difficult-interpretation
objects and blends in different boxes in order to better illustrate
the three samples (panels a, b and c).

\begin{figure*}
  \centering
  \includegraphics[clip=true, trim = 5mm 0mm 5mm 0mm, width=\textwidth, keepaspectratio]{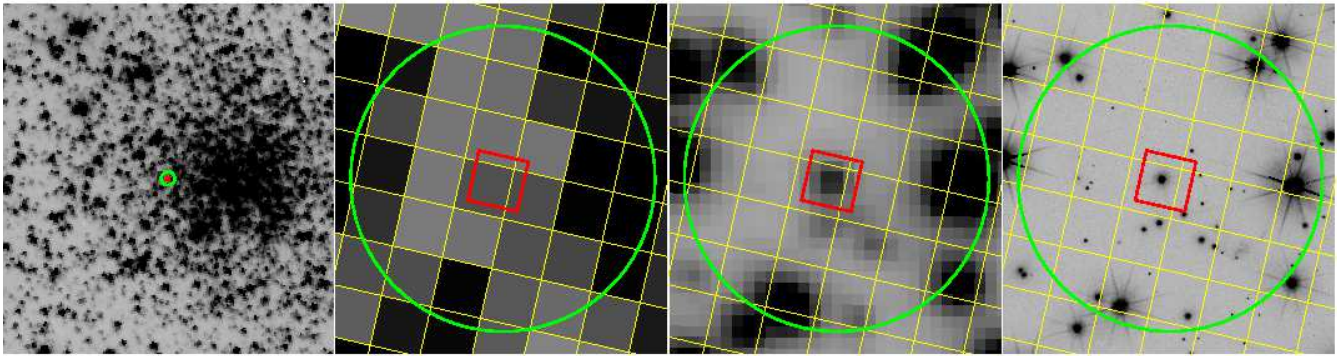}
  \caption{(\textit{Left}): 10$\times$10 arcmin$^2$ region around TR1
    in the \textit{K2} stacked image. A green circle of radius 3
    \textit{Kepler} pixels and a red square with 1 \textit{Kepler}
    pixel side are centered on TR1. (\textit{Middle-left}): zoom-in of
    the \textit{K2} stacked image around TR1. The covered area is
    about 6.5$\times$6.8 pixel$^2$ (about 25$\times$28
    arcsec$^2$). The yellow grid represents the \textit{Kepler} CCD
    pixel grid. (\textit{Middle-right}): as in the
    \textit{Middle-left} panel, but for the Schmidt filter-less
    stacked image of \citet{Nar15}. (\textit{Right}):
    ACS/WFC@\textit{HST} F606W-filter stacked image (from
    \citealt{Bed10}). In all these panels, North is up and East to the
    left. It is clear that the higher the image resolution, the higher
    the number of detectable polluting sources within the aperture.}
  \label{fig20}
\end{figure*}

\begin{figure*}
  \centering
  \includegraphics[width=\textwidth]{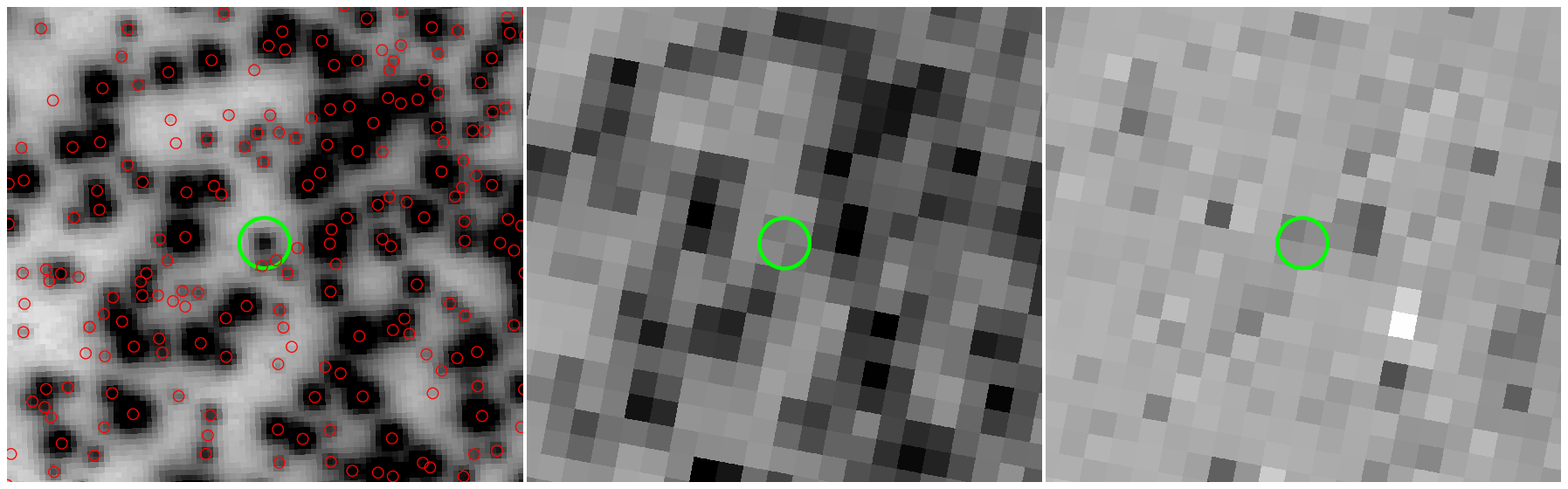}
  \caption{PSF-subtraction of the stellar neighbours (red circles)
    around TR1 (green circle). In the \textit{Left} panel, we show all
    identified sources in the AIC that must be subtracted before
    measuring the TR1 magnitude. The image is the Schmidt stacked image from
    which the AIC was extracted. In the \textit{Middle} and
    \textit{Right} panels we show a single \textit{K2} image before
    and after the neighbour subtraction, respectively. Even if the
    PSF-based subtraction is not perfect due to the unavailable PSF
    calibration data (see Sect.~\ref{PSF}), the light pollution is
    less severe.}
  \label{fig21}
\end{figure*}

\section{TR1 as a procedure benchmark}
\label{TR1sect}

\cite{Moc04,Moc06} made an extensive ground-based campaign search for
transiting exoplanets in NGC~2158 and, among the discovered variable
sources, they found an exoplanet candidate with a transit depth of
$\sim$0.037 magnitudes (TR1, following their
nomenclature). \cite{Moc06} suggested that TR1 could be a hot-Jupiter
with a period of 2.3629 days. The hosting star is a NGC~2158 member,
$V_{\rm max}\simeq$19.218, $R_{\rm max}\simeq$18.544,
$(\alpha,\delta)_{\rm J2000.0}\sim(\rm
06^h07^m35^s\!\!.4,+24^{\circ}05^{\prime}40^{\prime\prime}\!\!.8)$.

This object represents an ideal test-bed for our independent pipeline
reduction of \textit{Kepler}/\textit{K2} data in crowded
regions. Figure~\ref{fig20} shows the region of sky around TR1 in
images collected with three different instruments and completely
different resolutions. The light pollution of target neighbours is
evident.
In the left panel, we show an $\sim$10$\times$10 arcmin$^2$ region
(North up and East left) around TR1 from the \textit{K2} stacked image
of channel 81. A green circle of radius 3 \textit{Kepler} pixels and a
red square of 1 \textit{Kepler} pixel per side centered on TR1 are
barely visible in this panel.
In the middle-left panel, a zoomed-in image of about 25$\times$28
arcsec$^2$ centered on TR1. The yellow grid represents the
\textit{Kepler} CCD pixels. The red square shows the location of
TR1. It is clear that, without knowing the positions of the target,
TR1 identification would be hard if not impossible.
The middle-right panel shows the same region, but from the Schmidt
filter-less stacked image of \cite{Nar15}. In this image the pixel
scale is $\sim$0.862 arcsec pixel$^{-1}$. The higher spatial
resolution of this instrument allows us to better identify TR1. We can
identify at least 11 TR1 neighbours within the 3-\textit{Kepler}-pixel
aperture. These stars badly pollute the target LC, dimming the
transit, and therefore leading to an underestimated radius if we do
not properly account for their contribution. Finally, in the
right-most panel, we show the same region as seen in an
ACS/WFC@\textit{HST} F606W-filter stacked image described in
\cite{Bed10} (from GO-10500, PI: Bedin). In this case, the pixel scale
is about 25 mas pixel$^{-1}$ and the image shows that, within a single
\textit{Kepler} pixel, there can be more than one
star. Figure~\ref{fig21} shows an individual \textit{K2} exposure
before and after subtraction of neighbour sources present in the input
list.

The \textit{Kepler} magnitude of TR1 is $K_{\rm P,
  max}$$\simeq$18.35. Therefore, this object is at the faint end of
most of the previous studies. With the photometric technique developed
in this paper, TR1 becomes a well-measurable object, as we can push
our photometry almost 5 magnitudes fainter and measure stars in
crowded environments. In this magnitude interval and at this level of
crowding the 1-pixel-aperture and the PSF-based photometry are the
only two photometric approaches that allow us to measure the light
dimming in TR1 LC.

\begin{figure}
  \centering
  \includegraphics[width=\columnwidth]{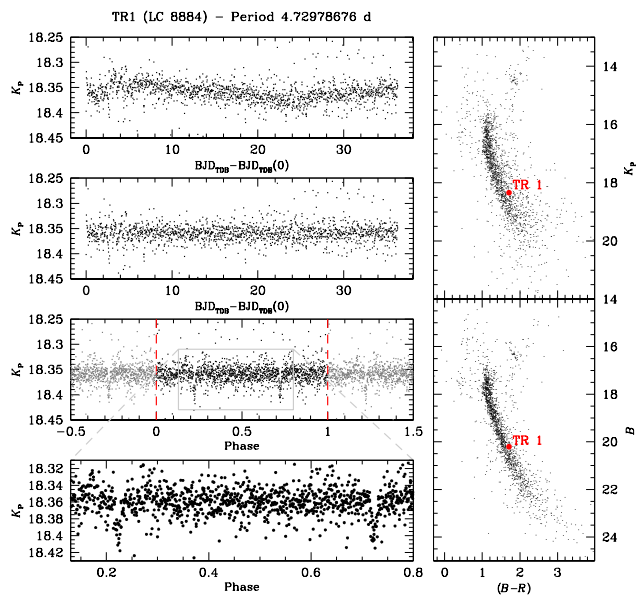}
  \caption{TR1 clean LCs and CMD location. (\textit{Left}): from
    \textit{Top} to \textit{Bottom}: detrendend LC, flattened LC (see
    text for details), phased LC with a period of 4.72978676 days and
    zoom-in on the phased LC to highlight the two minimums. In the
    one-to-the-last panel, the red, vertical dashed lines are set at
    phase 0 and 1. (\textit{Right}): $K_{\rm P}$ vs. ($B-R$)
    (\textit{Top} panel) and $B$ vs. ($B-R$) (\textit{Bottom} panel)
    CMDs of NGC~2158. We considered as cluster members those stars
    that are within a radius of $\sim$3 arcmin from NGC~2158
    center. The red filled point marks the location of TR1. The
    $K_{\rm P}$ magnitude was computed as the 3$\sigma$-clipped median
    value of the PSF-based LC magnitude, calibrated into the
    \textit{Kepler} photometric system (Sect~\ref{KPphot}).}
  \label{fig22}
\end{figure}

Despite the \textit{Kepler} pixel size, and consequent neighbour
contamination, the detrended light curve from \textit{K2} data is
significantly more complete and precise than that of \cite{Moc06}.
Figure~\ref{fig22} shows the phased and detrended clean LCs. In the
top panel, we plot the original, detrended LC. In the other panels we
show the same LC after we applied a running-average filter of 24-h
window to remove any systematic trend and/or long-period LC
variability in order to better highlight TR1 transits. The period we
found using the BLS algorithm is of 2.36489338 days. In
Fig.~\ref{fig22} we phased the light curve with double the
period.

In the phased LC, both eclipses have more a ``V''-like than a
``U''-like shape\footnote{As suggest by the referee, using
  \texttt{\small EXOFAST} tool (\citealt{EGA13}) we found an impact
  parameter of 0.764114 and a "planet" (star in TR1 case) to star
  radius ratio of 0.184497, confirming our qualitative
  classification.}. If we also consider the location of TR1 in the CMD
(\textit{Right} panel of Fig.~\ref{fig22}), it appears more likely to
be a binary with a grazing eclipse rather than a transiting exoplanet
candidate. In a future paper of this series we will make use of the
input list from \textit{HST} data to better characterize TR1. We will
also better determine the LCs for the subsample of stars that fall
within the ACS/WFC footprint of the \textit{HST} program GO-10500
field.

\begin{figure*}
  \centering
  \includegraphics[width=\textwidth]{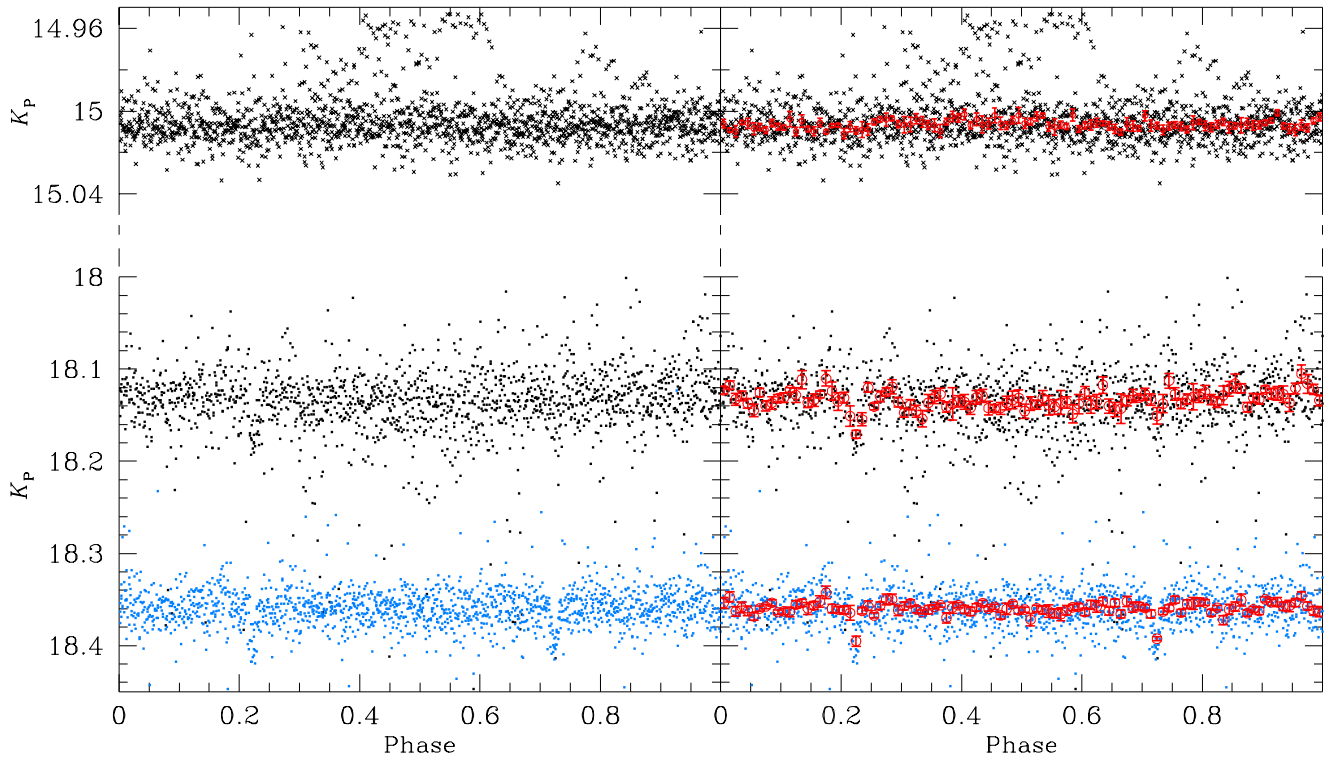}
  \caption{TR1 clean LCs with and without TR1 neighbours. From
    \textit{Top} to \textit{Bottom}: 3-pixel-aperture LC with
    neighbours (black crosses), PSF-based LC with (black dots) and
    without (azure dots) neighbours. The LCs are phased with a period
    of 4.72978676 d. In the \textit{Right} panels we also plot the
    binned LCs. The red points are the median value in 0.01-phase bin,
    while the error bars are the 68.27$^{\rm th}$ percentile of the
    distribution around the median value divided by $\sqrt{N}$, where
    $N$ is the number of points in each bin.}
  \label{fig23}
\end{figure*}

An important by-product of our method is that we can much-better
measure the depth of the eclipses, since our transits are less diluted
by the light coming from neighbour sources than other
approaches. Figure~\ref{fig23} shows a comparison between the
detrended clean TR1 LCs with (black points) and without (azure points)
TR1 neighbours. The uppermost LCs were those obtained with
3-pixel-aperture photometry, while those in the middle and in the
bottom were extracted using the PSF photometry. We phased these light
curves with a period of $\sim$4.73 d as in Fig.~\ref{fig22}. In the
right panel we also binned the LCs with bins of 0.01 phase each and
computed, within such each bin, the median and the 68.27$^{\rm th}$
percentile of the distribution around the median values.

The 3-pixel-aperture-based LCs do not show any significant flux
variation. The median magnitude is three magnitudes brighter and the
rms is smaller than in the PSF-based LCs; all evidences of
neighbour-light contamination as discussed in Sect.~\ref{photprec}.
This LC represents the result of a classic approach found in the
literature.

On the other hand, the remaining PSF-based LCs with (hereafter wLC)
and without (w/oLC) neighbours clearly present at least one flux
drop. First of all, TR1 in the wLC is about 0.25 magnitudes brighter
than in the w/oLC. The w/oLC does not only show a smaller rms and
fewer outliers, but it shows two distinct eclipses; while the wLC
exhibits only the eclipse around phase $\sim$0.2.

Using \texttt{\small VARTOOLS} BLS and \texttt{\small EXOFAST} suit
(\citealt{EGA13}, assuming TR1 is a transiting exoplanet), we
estimated a TR1 eclipse depth on the w/oLC (assuming the original BLS
period of $\sim$2.36 days) of about 3.2\% and 3.4\%,
respectively. Taking into account (1) that our LC and that of
\cite{Moc06} are obtained from different pass-bands, (2) the
measurement errors, (3) the \textit{K2} integration time lasts half an
hour, and (4) the incompleteness of the AIC, we conclude that the two
values are in rather good agreement. Therefore, we confirm that our
PSF-based approach is effective in disentangling blended sources in
crowded regions.

We conclude this section remarking that it would have been simply
impossible to extract the light curve of TR1 from
\textit{Kepler}/\textit{K2} data without using a input list from an
higher resolution data set and subtracting the stellar neighbours.
The approach of our method allows us to reach a 6.5-h photometric
precision of $\sim$2700 ppm in a heavily crowded environment for
sources as faint as TR1.

\section{Electronic material}
\label{electronic}

For each source in the AIC that fall in a \textit{K2}/C0/channel-81
TPF, we release raw and detrended light curves from the 1-, 3-, 5-,
10-pixel aperture and PSF photometry on the neighbour-subtracted
images. We also make public available the AIC (with each star flagged
as in/out any TPF) and the \textit{K2} stacked image.

For the 2849 candidate variables we
release\footnote{\href{http://groups.dfa.unipd.it/ESPG/Kepler-K2.html}{http://groups.dfa.unipd.it/ESPG/Kepler-K2.html}}
a catalogue made as follow. Column (1) contains the ID of the star in
the AIC. Columns (2) and (3) give J2000.0 equatorial coordinates in
decimal degrees; Columns (4) and (5) contain the pixel coordinates $x$
and $y$ from the AIC, respectively. In columns (6) and (7) we release
the instrumental Schmidt filter-less and the $K_{\rm P}$ magnitudes of
the stars. The $K_{\rm p}$ magnitude is computed as the
3$\sigma$-clipped median value of the magnitude in the LC, calibrated
using a photometric zero-point as described in Sect~\ref{KPphot}. In
column (8) we write the photometric method with which we extracted the
LC of the object and searched for variability.

Column (9) contains a flag corresponding to our by-eye classification
of the LC (Sect.~\ref{search} and \ref{literature}):
\\ $\bullet$ 0: high probability that it is a blend;
\\ $\bullet$ 1: candidate variable;
\\ $\bullet$ 2: difficult to classify;
\\ $\bullet$ 30: star marked as ``difficult to classify'' and that, by
comparison with the literature, could be a possible blend;
\\ $\bullet$ 31: star marked as ``difficult to classify'', but for
which we found a correspondence in the literature;
\\ $\bullet$ 32: star we classified as candidate variable but it is
close to a variable star from the literature and that seems a possible
blend.

Column (10) gives the period, when available. From column (11) to (16)
we give the ID used in other published catalogues, namely
\cite{Nar15}, \cite{Hu05}, \cite{Jeon10}, \cite{Kim04}, \cite{Meib09},
\cite{Moc04,Moc06}, respectively. Finally, we provide (columns from 17
to 22) the $B$$V$$R$$J_{\rm 2MASS}$$H_{\rm 2MASS}$$K_{\rm 2MASS}$
calibrated magnitudes from \cite{Nar15} catalogue, when available.

\section{Conclusions and Future planned works}

In this paper we have presented our first analysis of \textit{K2}
data, focusing our effort on crowded images and faint stars. The
test-beds for our method were super-stamps covering the OCs M\,35 and
NGC~2158.

Though the lack of \textit{Kepler} calibration data -- not made
available yet to the community -- prevented us to optimize our
algorithm, based on a technique we have developed in the last 20 years
on \textit{HST} undersampled images, we nevertheless succeeded in
implementing a photometric procedure based on the ePSF concept
\citep{AK00}. We have shown that by using a crude PSF that is
spatially constant across the channel and allowing for simple temporal
variations, we were able to properly fit stellar objects. Future
efforts will be devoted to further improve the PSF model, possibly
with a better data set at our disposal.

In the second part of the paper, we focused our attention on the
light-curve-extraction method and on the consequent detrending
algorithms. The LC extraction is based on the methods we started to
develop in \citet{Nar15} and makes use of both PSF fitting and of an
high-angular-resolution input list to subtract stellar neighbours
before measuring the stellar flux of any given target. By subtracting
the light of the close-by stars, we are able to decrease the dilution
effects that significantly impact the photometry in crowded regions.

We compared aperture- and PSF-based photometric methods and found that
aperture photometry performs better on isolated, bright stars
(6.5-hour-rms best value of $\sim$30 ppm), while PSF photometry shows
a considerable improvement with respect to the classical aperture
photometry on faint stars and in crowded regions. The extension of the
capability to exploit \textit{Kepler} as well as \textit{K2} data set
to fainter (up to 5 magnitudes fainter than what has been done up to
now in the literature) stars and crowded environments is the main and
original contribution of our efforts.

We release our raw and detrended LCs with the purpose of stimulating
the improvement of variable- and transit-search algorithms, as well as
of the detrending methods.

We are currently working on other clusters imaged during other
\textit{K2} Campaigns, and plan to work on the densest Galactic-bulge
regions. We also plan to go back to open clusters within the
\textit{Kepler} field, i.e. NGC~6791 and NGC~6819.

\section*{Acknowledgments}

We thank the referee Dr. A. Vanderburg for the careful reading and
suggestions that improved the quality of our paper. ML, LRB, DN and GP
acknowledge PRIN-INAF 2012 partial funding under the project entitled
``The M4 Core Project with Hubble Space Telescope''.  DN and GP also
acknowledge partial support by the Universit\`a degli Studi di Padova
Progetto di Ateneo CPDA141214 ``Towards understanding complex star
formation in Galactic globular clusters''. The authors warmly thank
Dr. Jay Anderson for the discussions and improvements to the
text. Finally, we thank Dr. B. J. Mochejska for the discussion about
TR1.

\appendix

\section{Example of light curves}
\label{appendixA}

\begin{figure*}
  \centering
  \includegraphics[width=0.8\textwidth]{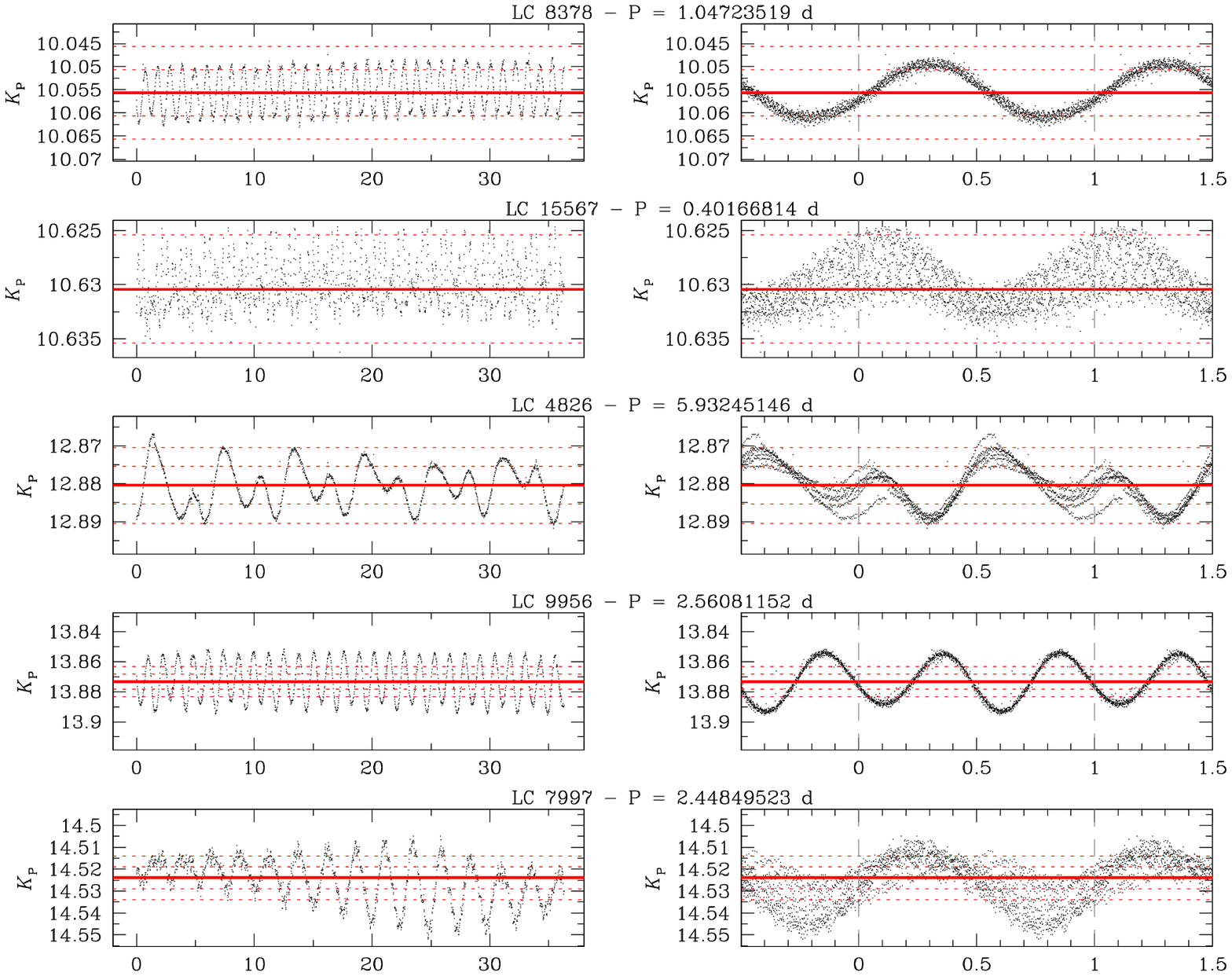}
  \vskip -5 pt
  \includegraphics[width=0.8\textwidth]{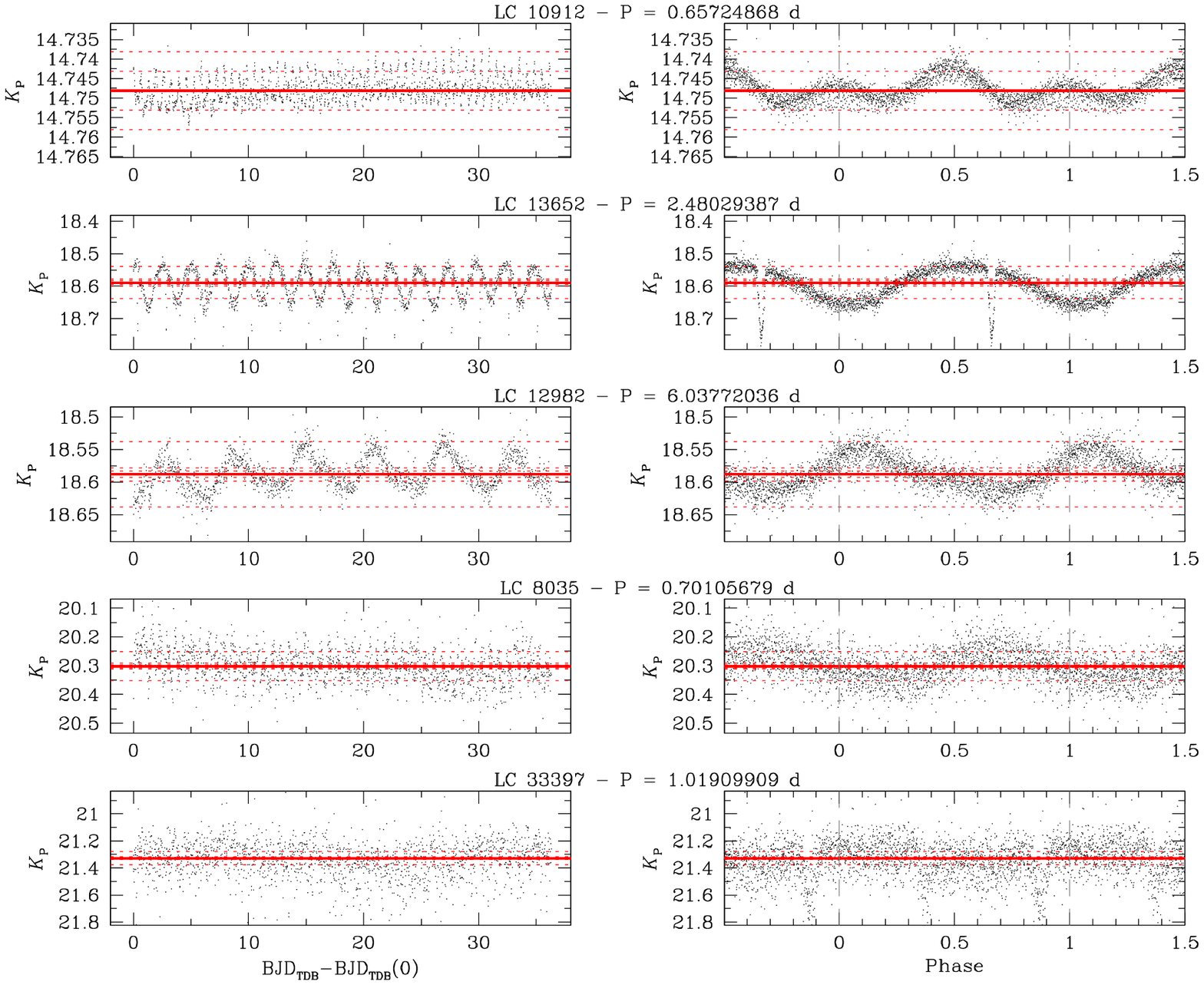}
  \caption{Example of variable light curves in our sample. We plot on
    the \textit{Left} and on the \textit{Right} the entire and the
    folded detrended clean LC of each object. The red solid line is
    set at the median magnitude of the LC, while the red dashed lines
    are set at $\pm$0.005, $\pm$0.01 and $\pm$0.05 $K_{\rm P}$ from
    the median value. The vertical, gray dashed lines in the
    \textit{Right} panels are set at phase 0 and 1. The LCs are
    plotted in order of decreasing magnitude from the top to the
    bottom.}
  \label{afig1}
\end{figure*}

In this appendix we show 10 of our light curves as example
(Fig.~\ref{afig1}).

\label{lastpage}

\end{document}